# Quantized Dislocations

Mingda Li[1]

[1]Department of Nuclear Science and Engineering, MIT, Cambridge, MA02139

## Abstract

A dislocation, just like a phonon, is a type of atomic lattice displacement but subject to an extra topological constraint. However, unlike the phonon which has been quantized for decades, the dislocation has long remained classical. This article is a comprehensive review of the recent progress on quantized dislocations, aka the "dislon" theory. Since the dislon utilizes quantum field theory to solve materials defects problems, we adopt a pedagogical approach to facilitate understanding for both materials science and condensed matter communities. After introducing a few preliminary concepts of dislocations, we focus on the necessity and pathways of dislocation's quantization in great detail, followed by the interaction mechanism between the dislon and materials electronic and phononic degrees of freedom. We emphasize the formality, the new phenomena, and the predictive power. Imagine the leap from classical lattice wave to quantized phonon; the dislon theory may open up vast opportunities to compute dislocated materials at a full quantum many-body level.

## 1. Introduction

A crystal dislocation, or dislocation for short, is a very common type of materials irregularity in crystalline solids [1,2]. This article is a comprehensive review of the recent theoretical progress on the quantization of the dislocation, resulting in a new quasiparticle: the "dislon". We will introduce the concept of the dislon, the necessity of quantization, and focus on the interaction mechanisms between dislons, electrons, and phonons, where new phenomena and new predictions arise. In this section, we first provide a self-contained overview on the background of a classical dislocation, including its equivalent definitions, its essential elements, and a few classical theoretical models and phenomenology related to functional properties.

### 1.1 The definitions of a dislocation

A clear understanding of a classical dislocation is the prerequisite to understand its quantization. There are at least three equivalent ways to define a dislocation, each with increased mathematical complexity but meanwhile increased feasibility toward quantization.

The first definition is shown in Figure 1a, which is also extensively introduced in materials science textbooks [3-5]. In this definition, a dislocation is classified as two prototypes – an edge dislocation (Figure 1a) and a screw dislocation (Figure 1b). Edge dislocation is like an insertion of atomic half-plane (orange atoms in Figure 1a), while screw dislocation is like a relative shift between two half-crystals (atoms on the left shift upward and atoms on the right shift downward in Figure 1b). Although straightforward to visualize, this type of dislocation is defined case-by-case on a particular lattice, hence is not suitable to develop a general quantized theory of dislocation.

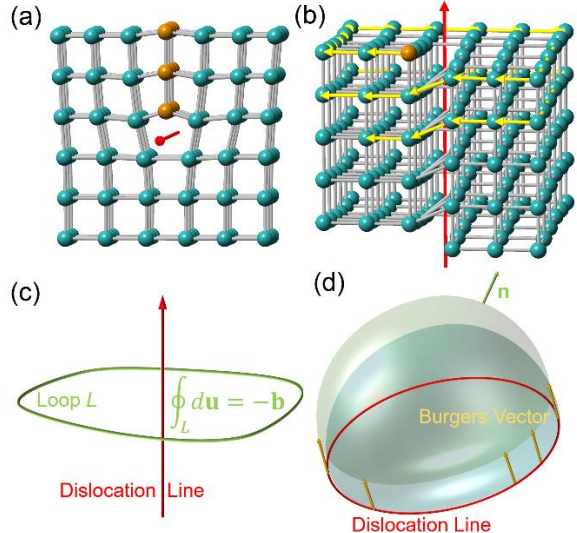

**Figure 1.** Three equivalent definitions of a dislocation. (a) Edge (left) vs screw (right) dislocation, where the helical "screw" shape surrounding the dislocation line is plotted with yellow arrows. (b) Dislocation defined through loop integral of displacement **u**. (c) Dislocation defined through creation of a discontinuity on a surface whose boundary gives the dislocation line. The dislocation lines are all drawn in red.

The second definition is rigorous, at least in the continuous limit. Mathematically, defining the lattice displacement **u** as the difference between the atomic position **R** in a dislocated crystal and the atomic position $\mathbf{R}_0$ in the corresponding pristine crystal, i.e. $\mathbf{u} \equiv \mathbf{R} - \mathbf{R}_0$, a dislocation can then be defined through a vector called the Burgers vector **b** [1]:

$$\oint_L d\mathbf{u} \equiv -\mathbf{b} \tag{1}$$

where $L$ is an arbitrary loop enclosing the dislocation line, called Burgers circuit (Figure 1b). Despite being highly generic, the arbitrary choice of the loop $L$ seems not quite helpful to lead to quantization either.

The third definition is also rigorous but originates from the theory of elasticity. Assuming that the $i^{th}$ component of lattice displacement $\mathbf{u}$ is $u_i$ ($i$=1,2,3, or equivalently $x, y, z$), the spatial derivative of $\mathbf{u}$ along the $j^{th}$ direction gives the so-called distortion tensor $\omega_{ij}$:

$$\omega_{ij} \equiv \partial u_i / \partial R_j \qquad (2)$$

The symmetrized version of $\omega_{ij}$ is called the strain tensor, $u_{ij} \equiv (\omega_{ij} + \omega_{ji})/2$. Then, a dislocation can be generated by creating a constant discontinuity of $u_{ij}$. For a given dislocation loop (red line in Figure 1d), we create an arbitrary surface bounded by the dislocation loop (blue-colored surface in Figure 1d). Then a dislocation can be created by a constant $\delta$-function–singularity on the surface, with an amount of the Burgers vector $\mathbf{b}$ along the surface normal direction $\mathbf{n}$ (green-colored surface in Figure 1d). If we define $\zeta$ as the local coordinate from the surface along the surface normal $\mathbf{n}$ direction (i.e., $\zeta$ is always locally perpendicular to the surface), then the above operation can be written as

$$\omega_{ij}(\mathbf{R}) = n_i b_j \delta(\zeta) \qquad (3)$$

It can be verified that Eq. (3) satisfies Eq. (1) by using Stokes' theorem [6]. The third definition, although cumbersome, actually facilitates the quantization procedure since it provides a feasible approach to perform a mode expansion on the displacement $\mathbf{u}$. In fact, the underlying elasticity theory that leads to the third definition of dislocation is quite general – it originates from a force equilibrium condition and can even be extended to other types of defects beyond dislocation, such as thin twinned layer [6].

### 1.2 The essence of a classical dislocation

With the knowledge of a dislocation's definition, we further introduce a few essential elements of a classical dislocation (Figure 2). From a perspective that is closely related to quantization, a classical dislocation contains at least three essential elements: "crystalline, topological, and dimensional".

• *Crystalline:* A dislocation only exists in crystalline solids where atoms are arranged in periodic order, but not in amorphous or liquid states. The crystalline requirement is critical for dislocation quantization. The applicability of Bloch's theorem in crystalline solids allows easy incorporation of the quantum theory of electrons and phonons, where crystal momentum $\mathbf{k}$ is good quantum number and can thus be used to label the corresponding quantum states. Even if the presence of dislocation often breaks the crystal symmetry, few approaches can restore the original symmetry, such as effective field theory and impurity average.

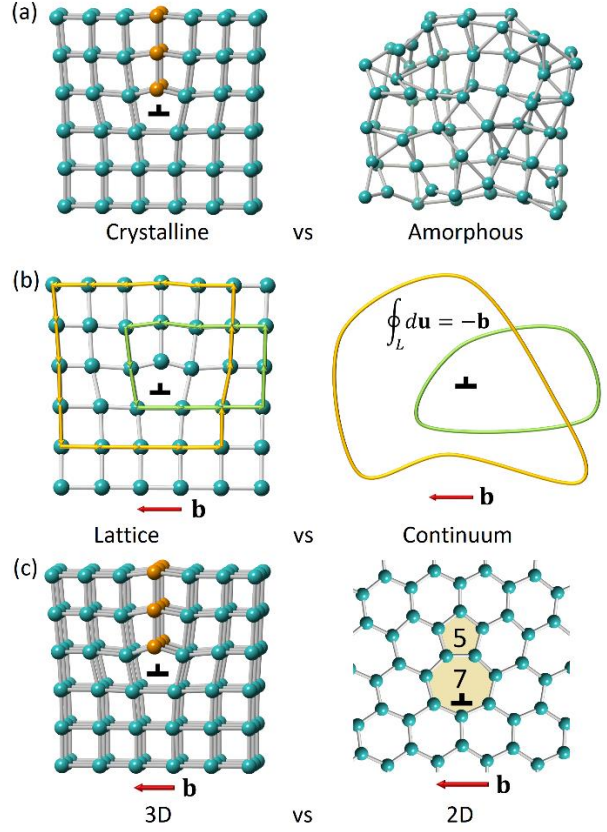

**Figure 2.** Essential elements of a dislocation: crystalline, topological and dimensional. (a) Dislocation only exists in crystalline solids but not in amorphous materials. (b) The total differential displacement surrounding arbitrary Burgers circuits (yellow or green) gives the same Burgers vector, valid for both lattice and continuous medium. (c) Dislocation can exist in both 3D and 2D, with different types of dislocations. In both 3D and 2D, a dislocation is well characterized by Burgers vector $\mathbf{b}$.

Although dislocations only exist in crystalline solids, the general definition of a dislocation, Eqs. (1) and (3), are also valid in a continuous medium. This validity greatly facilitates the construction of a long-wavelength, low-energy effective field theory. For instance, when studying electronic properties in a dislocated crystal, we can use Fermi liquid to model electrons without having to go through a lattice theory.

• *Topological:* A dislocation is known as a topological defect. Intuitively, the concept of "topological" indicates a level of global robustness against local perturbation. As a result, dislocations can only terminate at the crystal boundary, or form a self-closed loop, but cannot simply

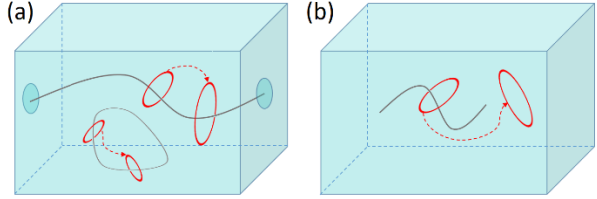

**Figure 3.** The topological invariance of a dislocation. For a dislocation line ending in crystal boundary (grey line in a) or forming a loop (grey loop in a), an arbitrary loop (red circle) surrounding the dislocation line will keep surrounding, no matter how it is continuously deformed. For a dislocation ending inside the bulk, shown in (b), a loop circling the dislocation can be continuously distorted into another loop without circling the dislocation anymore. This operation changes a finite **b** to **b**=0 and violates the topological invariance of dislocation.

start and end in the middle of the bulk (hence cannot be removed by local operations).

To understand why a dislocation forbids a "loose-end" inside the crystal, we take a closer look at Eq. (1): The arbitrariness of loop $L$ indicates that **b** is a topological invariant. In other words, any two loops continuously deformable to each other without touching the dislocation line shall give the same **b** (Figure 3a). However, dislocations ending inside the crystal bulk will inevitably change **b** from a finite value to **b**=0 (Figure 3b), contradicting the topological constraint definition Eq. (1). This topological constraint Eq. (1) is the main difference between a dislocation and a phonon, and shall be taken into account and respected in any quantized dislocation theory. Since both phonons and dislocations are intrinsically lattice displacements **u**, conceptually, we shall have

$$phonon = quantized\ \mathbf{u}$$
$$dislon = quantized\ \mathbf{u} + Eq.\ (1) \tag{4}$$

Regardless of a particular pathway of quantization, Eq. (4) can be considered as a philosophical starting point for any attempt to quantize a dislocation.

Another feature here is the physical implication of Eq. (1). Despite being seemingly innocuous, with the explicit appearance of displacement **u** only, Eq. (1) actually contains all scattering effects such as dynamic vibration, strain field, and Coulomb field, after quantization. An intuitive way to see this may come from a comparison to phonon: when quantizing the lattice displacement **u** of a phonon, the long-range electrical field generated by optical phonons can emerge naturally in polar materials [7].

• *Dimensional:* A dislocation is often referred to as a 1D line defect, aka "dislocation line". However, this is only valid in 3D solids; dislocations can also exist in 2D and quasi-2D systems, such as superconductor and superfluid thin films [8-11], membranes [12] and atomically thin 2D materials such as graphene [13-15]. Both dislocations in 3D and in 2D satisfy the definition Eq. (1) with well-defined Burgers vector **b**, but there are a few subtle differences. First, in 2D and quasi-2D, the dislocation line direction is always perpendicular to the 2D plane; a line defect lying within the 2D plane is often referred to as a grain boundary [16-18] (Figure 4a). Second, in 2D materials, the Burgers vector **b** also lies within the 2D plane, thus the common screw dislocation in 3D, where dislocation line is parallel to **b**, cannot exist in 2D [13]. Third, in 3D, another related defect called a disclination can hardly exist due to the very high energy, while in 2D, a disclination is more fundamental, and a dislocation can be considered as a pair of disclinations. Last, in 3D, the line defect of a dislocation does not resemble a point defect at all, while in 2D, dislocations and point defect clusters share some superficial structural similarity (Figure 4b, c).

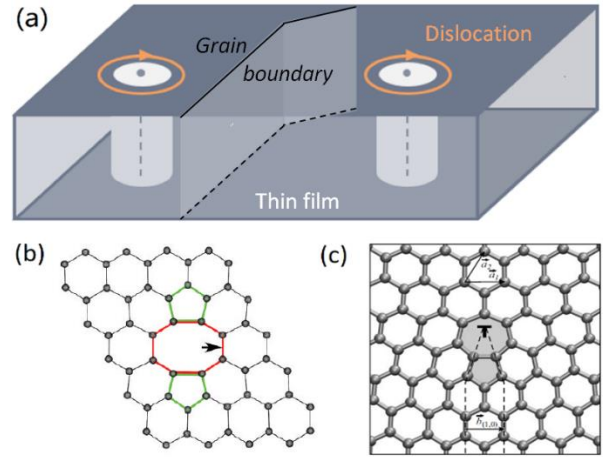

**Figure 4.** (a) Dislocation and grain boundary in a quasi-2D system. In 2D, a double vacancy pair 5-8-5 defect (b) and a dislocation 5-7 ring (c) share some superficial similarity. (b) and (c) are adapted from [13] and [15].

### 1.3 A brief history of dislocation

To gain a better understanding of dislocations, here we digress slightly and provide a crash-course on the early development of dislocations, with a few milestones listed in Figure 5.

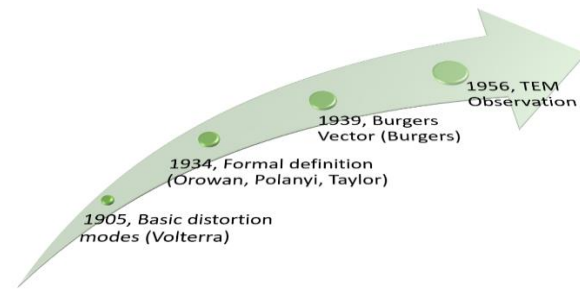

**Figure 5.** A few early milestones of dislocation.

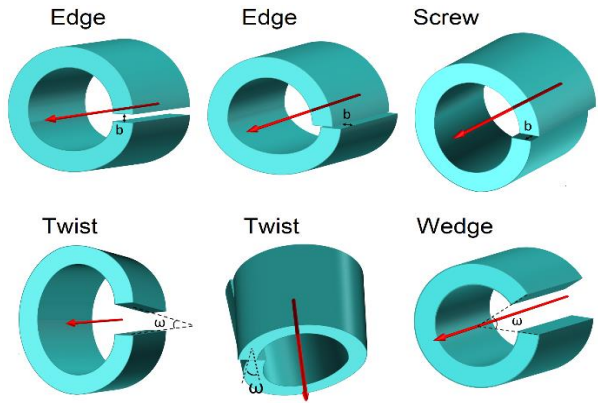

**Figure 6.** Six elementary deformation modes in a solid, including three types of dislocations (first row) and three types of disclinations (second row). The defect line directions are shown as red arrows, with corresponding Burgers vector and Frank vector labeled as **b** and **ω**, respectively.

• Volterra's construction: In 1905, V. Volterra published a series of theoretical papers discussing the elastic deformation modes in solids [19-24]. The key contribution is that he proved that all deformation in solids could be decomposed as a superposition of 6 basic modes- 3 are primitive dislocations while another 3 are primitive disclinations (Figure 6). The reason we can call them dislocations (or disclinations) is that these basic deformation modes will generate exactly the same stress-strain field distribution as real dislocations (or disclinations), at least away from the core region to avoid singularity. The reason why it is considered "primitive" only is due to the lack of the concept of crystal – as mentioned above, dislocation only exists in crystalline solid, yet the periodic crystalline structure was discovered by von Laue in 1912 using X-ray diffraction, seven years after Volterra's construction. What is interesting to mention is that in addition to the rigorous mathematical proof, Volterra also used plaster mold to build the edge and screw dislocations models in his original papers (Figure 7).

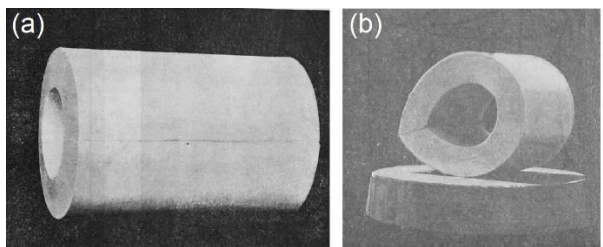

**Figure 7.** Prototypical deformation caused by primitive edge (a) and screw (b) dislocations, made of plaster mold. Figure adapted from Volterra's original publications [23,24].

• Formal definition: In 1934, E. Orowan, M. Polanyi, and G. I. Taylor theoretically invented the concept of dislocation, independently, with a level of mutual awareness of each other's work [25-27]. What they formulated was indeed an edge dislocation (Figure 8). Moreover, they applied the dislocation theory to study the materials plastic deformation process.

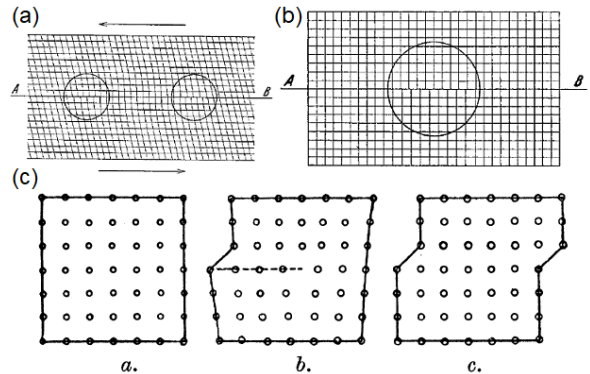

**Figure 8.** Edge dislocation plotted in Orowan (a), Polanyi (b) and Taylor (c)'s original publications.

• Burgers vector: The Burgers vector **b** was proposed by J. M. Burgers in 1939 [28,29]. It can be considered the central quantity that characterizes a dislocation. For instance, with the information of Burgers vector and dislocation line direction, the surrounding stress field distribution can be determined. The same information also determines the direction toward which a dislocation moves. Moreover, as a summable vector quantity, the Burgers vector also facilitates the consideration of multiple dislocations, such as dislocation dipole and dislocation networks.

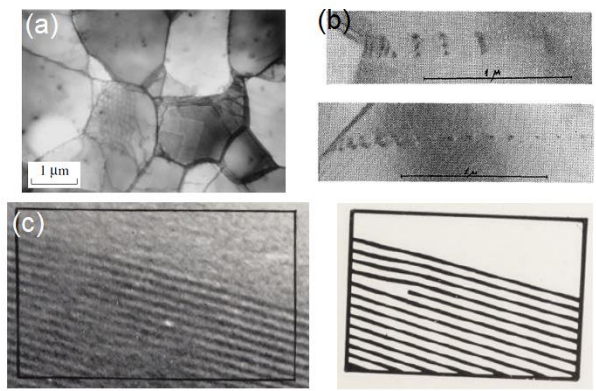

**Figure 9.** The TEM images of dislocations, by Hirsch in Al (a), Bollmann in steel (b) and Menter in PtPc (c). Figures adapted from [30-32]. In particular, (a) is generally considered as the first TEM image of dislocations.

• Direct observation: In 1956, thanks to the advancement of transmission electron microscopy (TEM), direct observation of dislocations was carried out by three groups of people, including Hirsch, Horne, and Whelan [33], Bollmann [31], and Menter [30] (Figure 9). The reason why the dislocation image in Menter's work (Figure 9c) looks different from Hirsch's (Figure 9a) and Bollmann's (Figure 9b) is due to the contrast mode. Hirsch and Bollmann used metals with small lattice parameters to see dislocations lines with diffraction contrast (i.e., single diffraction spot is used to reconstruct the image), Menter chose large unit-cell materials copper and platinum phthalocyanine (CuPc and PtPc), which enabled an inclusion of multiple diffraction spots in the aperture to form an image of the extra atomic plane of an edge dislocation, directly.

## 2 The Role of Dislocations on Functionalities

The glorious history and the rigorous theoretical framework of a classical dislocation may leave an impression that the problems associated with dislocations are solvable, at least in principle. Such an impression might be more-or-less true for mechanical properties. However, for non-mechanical, functional properties in a dislocated crystal, plenty of open questions remain to be answered, related to electronic structure, thermal transport, optical properties, magnetic ordering, and superconductivity. In this Section, we review some prominent theoretical models and experimental phenomena that are related to these functionalities. The examples are by no means complete, but only reflect the author's taste and limitation. Even so, we may be able to see some opportunities that can be created from these examples.

### 2.1 Electron-dislocation scattering and electrical transport

The electrical transport in dislocated materials has been extensively studied both theoretically [34-47] and experimentally [48-54] in both semiconductors [34,35,38-41,48,53,55] and metals [36,42-44,47,49-51,54]. A few most prominent studies share a common feature, that to develop a theoretical model first, followed by experimental measurements, with a focus on the carrier mobility and electrical resistivity [34,38-40,44,52,55].

For semiconductors, the electron mobility and its temperature dependence was calculated in a seminal study by Dexter and Seitz [34]. They obtained that the electron-dislocation scattering rate, $1/\tau_d$, has the following temperature dependence:

$$1/\tau_d = \frac{\alpha_d}{T} \tag{5}$$

where $\alpha_d$ is a constant. Also, the scattering potential $V(\mathbf{r})$ can be written as

$$V(\mathbf{r}) = -C \frac{b}{2\pi}\left(\frac{1-2\nu}{1-\nu}\right)\frac{\sin\theta}{r} \tag{6}$$

where $C$ is the so-called conduction-band deformation potential constant that is determined experimentally [34,56], $b$ is the Burgers vector length, $\nu$ is the Poisson ratio, $\theta$ is the angle relative to the direction of dislocation slip, and $r$ is the distance between the electron and the dislocation core.

What is worthwhile mentioning here is that the dislon theory can easily reproduce the results given in Eqs. (5) and (6), moreover determine the proportionality constant $C$. For temperature dependence, it can be shown that the $1/T$ dependence is a direct consequence of the lowest-order one-loop correction of electron self-energy $\Sigma$ [57],

$$1/\tau_d \propto \text{Im}\,\Sigma \propto \frac{1}{T} \tag{7}$$

with the following self-energy

$$\Sigma = \quad \text{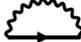} \tag{8}$$

in which the wavy line denotes the dislon propagator, while the straight line is the electron propagator.

Besides the relaxation time, the dislon theory also allows for a microscopic determination of the constant $C$ in Eq. (6) [58]:

$$C = \frac{2\pi n Z e^2}{k_{TF}^2} \tag{9}$$

where $n$ is the electron density in the crystal, $Z$ is the ionic charge number, $e$ is elementary charge, and $k_{TF}$ is the Thomas-Fermi screening wavevector.

A more recent prominent example of electron-dislocation scattering is related to GaN thin films for potential light emitting diode (LED) and high-power electronics applications [35,38,39,52]. Due to the lattice mismatch between the thin film and substrate, threading and misfit dislocations exist in GaN films (Figure 10). The threading dislocations can strongly scatter electrons and limit the mobility.

In this example, given the polar nature of the wurtzite III-V compounds and the weak Coulomb screening due to the low carrier density in semiconductors, the corresponding scattering mechanism is predominantly Coulomb scattering, where the dislocation is modeled as a line charge. The deficiency of this approach is apparent – by modeling a dislocation as a line charge, many other features of a genuine dislocation are missing. Even the validity of the line charge approximation is uncontrolled and unclear.

For metals, one may naturally expect that the deformation potential scattering Eq. (6) is the main mechanism that accounts for electron-dislocation interaction, yet the situation is complicated by the resonant scattering between electrons near the Fermi surface and the dislocation core [36,42-44,47]. One intuitive understanding of such resonance may arise from the comparable, sub-nm scale between the Fermi wavelength of the electrons and the dislocation core size [59]. Currently, the dislon theory has not been applied to the resonance electron-dislocation scattering, yet as a quantum field theory that is naturally capable of dealing with various resonant phenomena [60,61], the study of electron-dislocation interaction taking into account the resonance scattering would be straightforward. In particular, all the present studies treat the scattering process at a single-particle, first-order scattering level, and a field theory like the dislon theory would enable the incorporation of many-body effects such as electron-electron correlation and electron-phonon interaction in a unified manner.

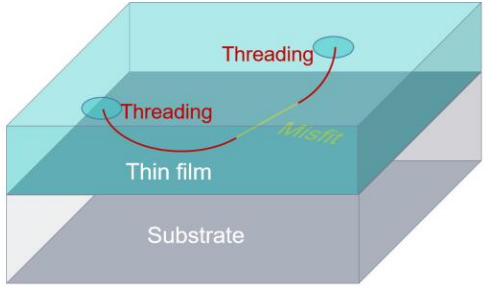

**Figure 10.** Threading and misfit dislocations in a thin film system. Threading dislocations connect the surface to the interface, while misfit dislocations lie along the interface.

### 2.2 Phonon-dislocation scattering and thermal transport

The nature of phonon-dislocation scattering is wildly different from that of electron-dislocation scattering. Starting from the 1950s till 1980s, there has been a three-decade-long debate regarding the mechanism of phonon-dislocation scattering, whether static strain scattering [62-68] or dynamic vibrational scattering [69-74]. The dynamic vibrational scattering is further divided into two subcategories, the drag-like fluttering mechanism [69-74], and the vibrational string mechanism [75-77]. All mechanisms contain numerous theories, such as the Klemens' and Carruthers' theories on static scattering [62,63], the Ninomiya's theory on fluttering [73], and the Granato-Lücke's vibrational string theory [75], with different temperature dependence of lattice thermal conductivity $k$:

$$k_{strain} \propto T^2$$
$$k_{flutter} \propto T^3 \qquad (10)$$
$$k_{string} \propto T^{7/2}$$

Although experimental reports that support the static mechanism do exist [68], it is generally believed that strain field is too small to induce a significant change of lattice thermal conductivity $k$. In particular, spanning the three decades, a series of measurements on the low-temperature thermal conductivity of LiF have been carefully performed (Figure 11) [78-85]. The wide electronic bandgap (13.6eV) [86], combined with the light elements hence high Debye frequency, makes LiF an ideal test platform for theories. As a result, a consensus was gradually reached that the dominant dislocation-phonon mechanism is coming from the fluttering mechanism [87].

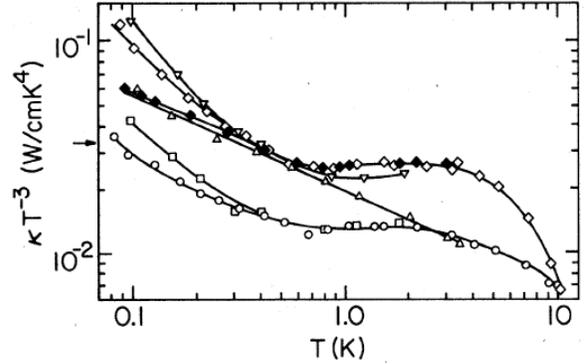

**Figure 11.** The low-temperature thermal conductivity of LiF down to 0.03K, on various samples with different deformation conditions. Figure adapted from [81].

The experimental-driven settlement of the debate left a few key questions unanswered. First, even if the static strain scattering is too weak to support experiments, its very existence is undeniable. It then becomes unclear when the strain field shall be taken into account. The existence of the strain field of dislocation also calls for a better theory that can treat both static scattering and dynamic vibration on equal footing. Second, there is also clear evidence showing the resonant dislocation-phonon scattering [88,89], which seems not properly explained in theory. Although the Granato-Lücke's vibrational string theory can create resonance modes, it is well-known that this theory is valid at low-frequencies only, such as explaining ultrasonic attenuation (~MHz range) [90], but not for phonons (~THz range) [81]. As we will see shortly in Section 5.1, by treating a dislocation as a quantum field, with both spatial extension and internal dynamics, the dislon theory is capable of solving the above questions directly.

### 2.3 Optical properties

In addition to electronic and thermal properties, dislocations also largely influence the optical properties. For instance, dislocations can either serve as "black holes" of charge carriers, aka non-radiative recombination centers, in semiconductors (Figure 12a) [91,92], or serve as "bleach" to decolor the irradiated alkali halide crystals when interacting with the color centers [93], along with other phenomena such as optical absorption and excitation [94-109]. One particularly promising phenomenon is the dislocation luminescence. When non-radiative recombination is suppressed, light emission can be achieved through a radiative recombination process, with potential LED applications [102-109]. At low temperature, four photoluminescence (PL) bands exist in dislocated Si, which are labeled as D1-D4 (Figure 12b) [102], while at room temperature, only D1 remains even after sample treatment to reduce non-radiative processes (Figure 12c) [103].

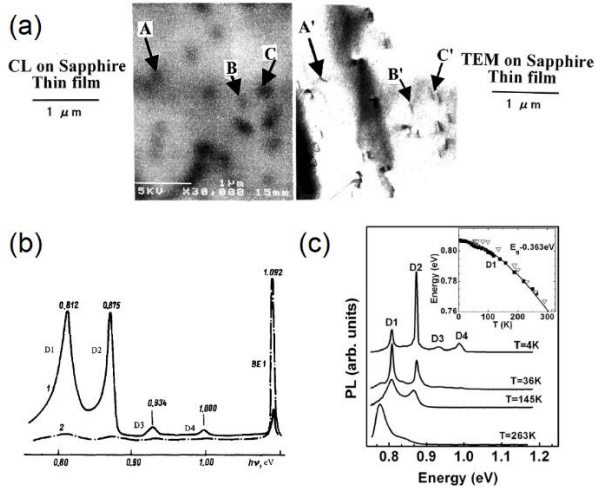

**Figure 12.** Non-radiative and radiative recombination processes of dislocations. (a) Dislocations as non-radiative recombination centers, where the dark spots on the cathodoluminescence (CL) spectra (left figure) match exactly the position of dislocations from TEM image (right figure). (b) Dislocation D-band luminescence at low temperature, labeled as D1-D4. (c) The D1 luminescence remains at room temperature after suppression of non-radiative carrier recombination. Figures are adapted from [91,102,103].

What is worth mentioning is that despite the extensive research on the dislocation's D-band luminescence, with numerous knowledge accumulated since the first observation in 1976 [102], the microscopic origin of these bands is still not fully understood, particularly for the D2 band [108-111]. In fact, research on optical properties of dislocated crystals is still largely driven by experiments, leaving large room for further theoretical investigation.

## 2.4 Magnetic ordering

Dislocations are also known to play a role in magnetic materials [112-128]. One primary mechanism that has been extensively studied is the interaction between dislocations and magnetic domain walls [123-128], where the strain field of dislocations affects the motion of the domain walls and thereby the magnetic hysteresis.

However, it appears that the interaction of dislocations with smaller objects, such as individual spins, is seldom reported. In one recent example, ferromagnetic-ordered dislocations are observed in antiferromagnetic material NiO, caused by Ni vacancies near the dislocation core region (Figure 13) [122]. In light of this, a microscopic understanding of dislocation-spin interaction may shed light on utilizing dislocations as a useful tool to tune magnetic phase transitions.

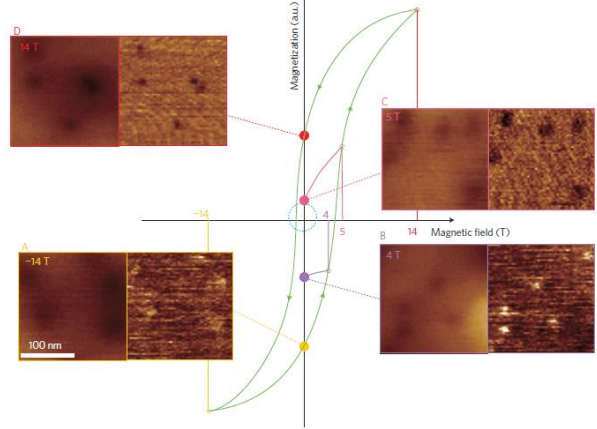

**Figure 13.** Ferromagnetic dislocations in antiferromagnetic materials. The magnetic hysteresis of dislocations in NiO, where the atomic force microscopy (AFM) topography and the magnetic force microscopy (MFM) are plotted on the left and right, respectively, at each external magnetic field. Figure adapted from [122].

## 2.5 Superconductivity

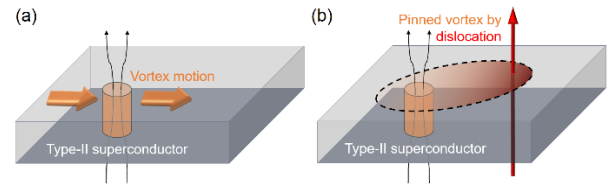

**Figure 14.** Illustration of flux pinning mechanism. (a) The motion of a quantized magnetic flux in a type-II superconductor. (b) The presence of dislocations can freeze the motion of the magnetic flux and increase the critical current and critical fields.

For superconductivity, dislocations are known to affect both the critical current $J_c$ and the transition temperature $T_c$. For critical current $J_c$, the mechanism is well established as the magnetic flux pinning, which has been studied extensively experimentally [129-140], such as in prototypical YBCO, and theoretically [141-151], such as using the Ginzburg-Landau theory. When a type-II superconductor is placed under strong magnetic field between first critical field $H_{c1}$ and second critical field $H_{c2}$, quantized vortices of magnetic flux may form with flux quantum $\Phi_0 = h/2e$. The flux motion, such as thermally activated flux creep, will create a pseudo-resistance and decrease the critical current $J_c$ (Figure 14a). In this sense, dislocations may behave as pinning centers that prevent the motion of the flux and thereby increase $J_c$ (Figure 14b).

The influence on transition temperature $T_c$ seems more involved. Although it is well known that dislocations can change the superconducting transition temperature [152-170], the exact mechanism is unclear, even in simple Bardeen-Schrieffer-Cooper (BCS) superconductors (Table 1).

**Table 1.** Superconducting transition temperature $T_c$ in pure and dislocated simple metals. Table adapted from [58].

| $T_c$ [K] | Al | In | Nb | Pb | Sn |
|---|---|---|---|---|---|
| Pure | 1.2 | 3.37 | 9.42 | 7.21 | 3.72 |
| Dislocated | 1.2 | 3.51 | 9.94 | 7.21 | 3.95 |
| | Ta | Ti | Tl | V | Zn |
| Pure | 4.46 | 0.49 | 2.2 | 5.47 | 0.9 |
| Dislocated | 4.46 | 0.37 | 2.48 | 5.94 | 1.39 |

The anisotropic superconducting gap [156-159] offers a possible theoretical foundation to explain $T_c$ with the presence of generic impurities [156-158]:

$$\frac{T_c - T_c^0}{\eta} = a + b \log \eta \qquad (11)$$

where $T_c^0$ is the transition temperature of the pure case, $\eta$ is the ratio between residual resistivity $\rho_{0K}$ and phonon-limited resistivity (say, room-temperature $\rho_{RT}$), i.e. $\eta = \rho_{0K}/\rho_{RT}$, $a$ and $b$ are empirical parameters. However, problems remain. First, the original theories are derived for generic quenched impurities [156-158] with an elastic scattering impurity potential, but not for dislocations which have internal dynamics. Second, it is still challenging to obtain a theory using neither empirical parameters nor experimental parameters to compute $T_c$. Third, different superconductors respond differently to dislocation density. For Al, the $T_c$ change is as small as few mK level even with high dislocation density [159], while for Zn, the enhancement can be huge. As we will see in Section 4.3, the dislon theory can conquer these shortcomings and explain $T_c$ at a quantitative level without free parameter.

### 2.6 Topological materials

Topological materials, such as topological insulators [171-173] and topological semimetals [174-176], are news categories of emergent condensed matter phases, where the topology of the bulk electronic bandstructures induces exotic electronic states at materials surface. Since the discovery of 2D topological insulator in the HgTe/CdTe quantum well [177,178], the 3D topological insulators in the $Bi_2Se_3$ family [179-182], and most recently the topological Weyl and Dirac semimetals in the TaAs family [183,184], many research efforts have been focused on searching for novel topological materials families, with a certain level of awareness of dislocations. In topological insulators, dislocations can provide dissipationless conduction channels [185-187], while in topological semimetals, dislocations may induce chiral magnetic effect [188,189] or emergent magnetic flux [190]. Moreover, dislocations can also serve as a platform to host other topologically protected modes [191-196]. For instance, Majorana fermions may be created at an edge dislocation core in a topological superconductor [192]. In these examples, instead of placing dislocations in the solid bulk and studying their interaction mechanism, dislocations play the role of local hosts that accommodate exotic states which do not exist in the solid bulk (Figure 15) [197].

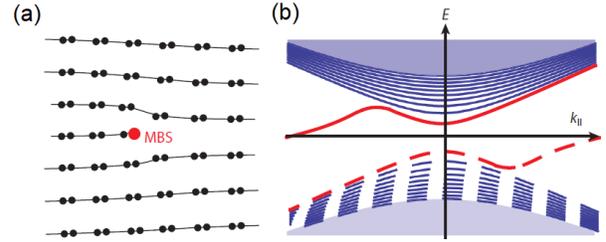

**Figure 15.** (a) Illustration of zero-energy Majorana bound state (MBS) at the core of an edge dislocation in a weak topological superconductor with first Chern number $C_1=0$. (b) Chiral Majorana edge modes (in red) of a strong topological superconductor with $C_1=1$. Figure adapted from [197].

## 3 Dislon as Quantized Dislocation

With the background knowledge of classical dislocations introduced, we are ready to introduce the quantized dislocation, aka the dislon, formally. We first provide a conceptual approach of quantization (Section 3.1), and then introduce the formalism for a classical dislocation (Section 3.2) and the corresponding quantization at the first-quantized level (Sections 3.3 - 3.5). After clarifying a few details (Section 3.6, 3.7), we will transit from the first-quantized dislocation to a second-quantized dislocation, aka the dislon (Section 3.8). The dislon Hamiltonian is derived (Section 3.10), with a few generalization schemes introduced at the end for more involved problems.

### 3.1 Classical to quantized dislocation: Concepts

In Section 1.2 and Eq. (4), we mentioned that in order to transit from a classical dislocation to a quantized one, we need to respect the Eq. (1). However, "respect" does not mean "satisfy". In fact, quantum fluctuations of the displacement, which account for all dynamic scattering effects such as fluttering phonon-dislocation, indicates that Eq. (1) must break down (but still be respected) whenever dynamic effect is considered. To gain an intuitive understanding, we refer to Figure 16.

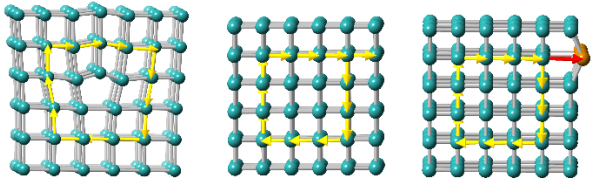

**Figure 16.** The breakdown of the dislocation's definition due to dynamic effect which contributes extra atomic displacement.

As commonly introduced in elementary Materials Science textbooks, a dislocation is defined through the Burgers circuit: a closed loop in a dislocated crystal becomes an incomplete loop when moving the same steps along each direction; the amount of failure of loop closure gives the topological invariant of the Burgers vector **b** (Figure 16, left and middle). However, when dynamical effect occurs (exaggerated as orange atoms in Figure 16, right), the loop-closure failure is no longer a constant but depends on the amount of local fluctuation. Moreover, the amount of closure failure further depends on where the loop is selected. This prevents the definition of a topologically-invariant Burgers vector. In short, Eq. (1) no longer holds with the presence of dynamical quantum fluctuation, but should still be respected as a classical static limit, or say boundary condition.

Putting into the conceptualized equation, we have

$$\text{Dislon} = \text{Quantum fluctuation} + \text{Classical dislocation} \tag{12}$$

Or more rigorously, Eq. (12) can be written as

$$\text{Dislon} = \text{Dynamic operator} + \text{Boundary condition} \tag{13}$$

The importance of the dynamic effect in a dislocation can hardly be overemphasized. It is a natural consequence of the internal structure of dislocations, since the spatial extension beyond quenched point defects necessitates its internal dynamical structures. Such dynamical structure may play a role in the resonance scattering and Coulomb scattering with electrons (Section 2.1), fluttering and vibrational string scattering with phonons (Section 2.2), dislocation induced PL due to induced energy levels (Section 2.3), and superconducting transition temperature $T_c$ (Section 2.5). Most importantly, by writing a dislocation in Eq. (13) and recognizing the quantum dynamic effect, it provides a systematic approach to study all phenomena induced by dislocations using one unified language, with classical effect already incorporated but without having to develop individual models for each functionality.

## 3.2 Mode expansion of classical dislocation

With the conceptual definition Eq. (13) in mind, we can formally construct a general theory to see what a quantized dislocation may look like. From Sections 1.2 and 1.3, we learned that in 3D, the dislocation line direction combined with the Burgers vector direction could fully determine the dislocation characteristics, including the displacement field. The displacement field of a classical dislocation, $\mathbf{u}_{cl}(\mathbf{R})$, can be generally written as:

$$\mathbf{u}_{cl}(\mathbf{R}) = \frac{1}{L^2} \sum_{\mathbf{k}} e^{i\mathbf{k} \cdot \mathbf{R}} \mathbf{F}(\mathbf{k}) \tag{14}$$

where $\mathbf{R}$ is the 3D spatial coordinate $\mathbf{R}$, $\mathbf{F}(\mathbf{k})$ is a vector expansion function that makes Eq. (14) satisfy Eq. (1), i.e. $\oint_L d\mathbf{u}_{cl} \equiv -\mathbf{b}$, and we assume that the dislocation lies inside a box with edge length $L$, with dislocation core location at $x_0 = y_0 = 0$; the inverse area $1/L^2$ prefactor is for later convenience.

Eq. (14) is generic and shall be valid for all type of dislocations, including the prototypical edge and screw dislocations. It also contains information about the dislocation type in the feature of $\mathbf{F}(\mathbf{k})$. Assuming we have a straight dislocation line along the z-direction, defining $\mathbf{R} = (\mathbf{r}, z) = (x, y, z)$, $\mathbf{k} \equiv (\mathbf{k}_{\parallel}, k_z) = (k_x, k_y, k_z)$, Eq. (14) can be rewritten as:

$$\mathbf{u}_{cl}(\mathbf{R}) \equiv \mathbf{u}_{cl}(\mathbf{r}, z) = \frac{1}{L^2} \sum_{\mathbf{k}} e^{i\mathbf{k}_{\parallel} \cdot \mathbf{r} + ik_z z} \mathbf{F}(\mathbf{k}_{\parallel}, k_z) \tag{15}$$

Since a dislocation in 3D is a line defect, for a straight line dislocation without dynamic effect (see in Figure 1a, b), there is unbroken translational symmetry along the dislocation line z-direction. Therefore, the expansion coefficient $\mathbf{F}(\mathbf{k}_{\parallel}, k_z)$ shall not have any explicit dependence on $k_z$ for a classical dislocation, i.e. $k_z = 0$. Defining $\mathbf{F}(\mathbf{k}_{\parallel}) \equiv \mathbf{F}(\mathbf{k}_{\parallel}, k_z = 0)$, the final expansion for a straight classical dislocation line along the z-direction can finally be written as

$$\mathbf{u}_{cl}(\mathbf{R}) = \frac{1}{L^2} \sum_{\mathbf{k}} e^{i\mathbf{k}_{\parallel} \cdot \mathbf{r} + ik_z z} \mathbf{F}(\mathbf{k}_{\parallel}) \tag{16}$$

Then, for an edge dislocation, $\mathbf{b} \perp z$, using Eq. (1) and Eq. (15), we have

$$\mathbf{F}_z(\mathbf{k}_{\parallel}) = 0, \text{ edge dislocation} \tag{17}$$

Similarly, for a screw dislocation along the z-direction, $\mathbf{b} \parallel z$, we have

$$\mathbf{F}_x(\mathbf{k}_\parallel) = 0 = \mathbf{F}_y(\mathbf{k}_\parallel), \text{ screw dislocation} \qquad (18)$$

For a more general form of expansion coefficient $\mathbf{F}(\mathbf{k}_\parallel)$, we will be able to describe a generic dislocation type with arbitrary $\mathbf{b}$ relative to the dislocation line $z$-direction. In other words, the mode expansion formalism, even for a classical dislocation, is not limited to any specific dislocation type, but can be applied to all straight dislocations.

### 3.3 Canonical quantization

To see how quantization can come into play, we assume a mode expansion of the lattice displacement:

$$\mathbf{u}(\mathbf{R}) = \frac{1}{L^2} \sum_{\mathbf{k}} e^{i\mathbf{k}\cdot\mathbf{R}} \mathbf{U}_{\mathbf{k}} \qquad (19)$$

where $\mathbf{U}_{\mathbf{k}}$ is the vector expansion that accounts for all non-classical effects beyond the classical static stress-strain field. Then, if we recognize $\mathbf{U}_{\mathbf{k}}$ as the canonical coordinate, the canonical conjugate momentum $\mathbf{P}_{\mathbf{k}}$ can be defined through a given Lagrangian $\mathcal{L}$, i.e. $\mathbf{P}_{\mathbf{k}} = \partial \mathcal{L}/\partial \dot{\mathbf{U}}_{\mathbf{k}}$, Then by naively imposing a canonical quantization condition

$$[\mathbf{U}_{\mathbf{k}i}, \mathbf{P}_{-\mathbf{q}j}] = i\hbar \delta_{\mathbf{k}\mathbf{q}} \delta_{ij} \qquad (20)\text{x}$$

where $i,j=$1,2,3, the quantization procedure is complete, where the expansion of the position $\mathbf{U}_{\mathbf{k}}$ and its conjugate momentum $\mathbf{P}_{\mathbf{k}}$ are promoted from functions in classical physics to the first-quantized operators in quantum mechanics.

### 3.4 Classical vs quantized dislocation: Formalism

If Eqs. (19) and (20) are the only starting point arising from a plane-wave expansion, it becomes identical to phonons. The key difference that distinguishes a dislon from a phonon is the boundary condition, written in Eq. (16): in the static limit without any dynamic fluctuation, the displacement field $\mathbf{u}(\mathbf{R})$ shall be reduced to a full classical displacement $\mathbf{u}_{cl}(\mathbf{R})$, i.e. we want to find a way to link $\mathbf{U}_{\mathbf{k}}$ to the classical boundary condition $\mathbf{F}(\mathbf{k}_\parallel)$. This has been carried out in detail in one recent dislon study [198]. Briefly speaking, we perform an inverse Fourier transform of Eq. (19), then

$$\mathbf{U}_{\mathbf{k}} = \frac{1}{L} \int d\mathbf{R} e^{-i\mathbf{k}\cdot\mathbf{R}} \mathbf{u}(\mathbf{R}) \qquad (21)$$

Then, the static classical limit of $\mathbf{U}_{\mathbf{k}}$ corresponds to the classical dislocation's displacement in Eq. (16):

$$\mathbf{U}_{\mathbf{k},cl} = \frac{1}{L} \int d\mathbf{R} e^{-i\mathbf{k}\cdot\mathbf{R}} \mathbf{u}_{cl}(\mathbf{R}) = \mathbf{F}(\mathbf{k}_\parallel) \delta_{0,k_z} \qquad (22)$$

where we have used the fact that $\int dz e^{-ik_z z} = L\delta_{k_z,0}$ for box normalization. Recognizing Eq. (22) as a boundary condition that the displacement expansion coefficient $\mathbf{U}_{\mathbf{k}}$ should satisfy, we can rewrite the boundary condition as

$$\lim_{k_z \to 0} \mathbf{U}_{\mathbf{k}} = \mathbf{F}(\mathbf{k}_\parallel), \text{ for } \forall \mathbf{k}_\parallel \qquad (23)$$

Eqs. (19), (20) and (23) thus form a rigorous definition born from the conceptualized definition of the dislon in Eq. (13), written in the first-quantized form.

If we review the dislocation's functional properties (Sections 2.1-2.6), we can see that all dynamic, quantum, internal energy effects can be incorporated into the dynamic operator $\mathbf{U}_{\mathbf{k}}$. This also indicates a possible generalization to a system-specific, multiple dynamic modes from Eq. (19),

$$\mathbf{u}(\mathbf{R}) = \frac{1}{L^2} \sum_{\mathbf{k}\lambda} e^{i\mathbf{k}\cdot\mathbf{R}} \mathbf{U}_{\mathbf{k}\lambda} \qquad (24)$$

where $\lambda$ is the label for the particular dynamic mode in a dislon.

### 3.5 The dislon formalism: *From vector to scalar quantization*

To further simplify the formalism, we recall the case of a phonon. The displacement of a phonon $\mathbf{u}^{ph}(\mathbf{R})$ can be written as [199]

$$\mathbf{u}^{ph}(\mathbf{R}) = \frac{1}{L^{3/2}} \sum_{\mathbf{k}} e^{i\mathbf{k}\cdot\mathbf{R}} \mathbf{U}_{\mathbf{k}}^{ph} \qquad (25)$$

where the mode expansion coefficient $\mathbf{U}_{\mathbf{k}}^{ph}$ is a vector operator. However, the three components of $\mathbf{U}_{\mathbf{k}}^{ph}$ are not independent, but are linked by the polarization vector $\boldsymbol{\xi}_{\mathbf{k}}$:

$$\mathbf{U}_{\mathbf{k}}^{ph} = u_{\mathbf{k}}^{ph} \boldsymbol{\xi}_{\mathbf{k}} \qquad (26)$$

in which $u_{\mathbf{k}}^{ph}$ is a scalar operator for the phonon mode expansion.

Here, for dislocation's dynamic effect, a natural choice that can reduce a vector operator to a scalar operator is

$$\mathbf{U}_{\mathbf{k}} = u_{\mathbf{k}} \mathbf{F}(\mathbf{k}) \qquad (27)$$

This reduction indicates that the fluctuation along different directions of a straight dislocation are not independent, but are linked by the function $\mathbf{F}(\mathbf{k})$ in Eq. (14). This is a reasonable statement, since it indicates that a larger displacement will ensure a larger fluctuation, but a more rigorous proof would be highly desirable. Combining Eq. (23) with Eq. (27), the vector boundary condition Eq. (23) can be reduced to the following scalar boundary condition

$$\lim_{k_z \to 0} u_k = 1 \qquad (28)$$

Eq. (28) is important. In the vector boundary condition Eq. (23), the vector operator $\mathbf{U_k}$ and the material-dependent expansion coefficient $\mathbf{F(k)}$ are coupled. When Eq. (27) is imposed, the boundary condition of the scalar operator $u_k$, Eq. (28), becomes material-independent, where the material-dependent coefficient $\mathbf{F(k)}$ is decoupled from the dynamic operator $u_k$. In either case, Eq. (23) or Eq. (28) will ensure that the all classical dislocation effects are always adequately taken into account.

### 3.6 The dislon formalism: *Construction of expansion coefficient* $\mathbf{F(k)}$

Up to this step, we have demonstrated the feasibility of quantizing a dislocation at the first-quantization level while respecting the classical dislocation as a boundary condition. However, aside from the general symmetry consideration of $\mathbf{F(k)}$ in Eqs. (17) and (18), the explicit expression of $\mathbf{F(k)}$ has not been considered. For a straight dislocation line along the $z$-direction, with the $xz$ plane taken as slip-plane, an explicit form of $\mathbf{F(k)}$ in an isotropic medium has been found by Ninomiya [73,74,200,201] when studying the fluttering effect:

$$\mathbf{F(k)} = \frac{\mathbf{n}(\mathbf{b \cdot k}) + \mathbf{b}(\mathbf{n \cdot k}) - \dfrac{\mathbf{k}(\mathbf{n \cdot k})(\mathbf{b \cdot k})}{(1-\nu)k^2}}{k_x k^2} \qquad (29)$$

where $\mathbf{n}$ is the slip-plane normal direction, and $\nu$ is the Poisson ratio. As a result, the displacement field of a quantized dislocation in the first-quantized form can be written as

$$\mathbf{u(R)} = \frac{1}{L^2} \sum_k e^{i\mathbf{k \cdot R}} \mathbf{F(k)} u_k \qquad (30)$$

where $\mathbf{F(k)}$ takes the form in Eq. (29) for an isotropic material, and $u_k$ is subject to the boundary condition Eq. (28). Instead of rederiving the form of $\mathbf{F(k)}$ from scratch, here we verify that Eq. (29) is indeed reducible to very familiar results for edge and screw dislocations.

For an edge dislocation along the $z$-direction with Burgers vector $\mathbf{b} = b\hat{x}$ and slip-plane normal $\mathbf{n} = \hat{y}$, we have

$$\mathbf{F}_x(\mathbf{k}_\parallel) = +\frac{b}{k_x k_\parallel^2}\left(k_y - \frac{1}{(1-\nu)}\frac{k_x^2 k_y}{k_\parallel^2}\right)$$

$$\mathbf{F}_y(\mathbf{k}_\parallel) = +\frac{b}{k_x k_\parallel^2}\left(k_x - \frac{1}{(1-\nu)}\frac{k_y^2 k_x}{k_\parallel^2}\right) \qquad (31)$$

$$\mathbf{F}_z(\mathbf{k}_\parallel) = 0$$

where $k_\parallel^2 \equiv |\mathbf{k}_\parallel^2|$. Substituting Eq. (31) to Eq. (16), we have

$$\mathbf{u}_{x,cl}(\mathbf{R}) = \frac{b}{2\pi}\left[\tan^{-1}\left(\frac{y}{x}\right) + \frac{1}{2(1-\nu)}\frac{xy}{r^2}\right]$$

$$\mathbf{u}_{y,cl}(\mathbf{R}) = -\frac{b}{2\pi}\left[\frac{1-2\nu}{2(1-\nu)}\ln r + \frac{1}{2(1-\nu)}\frac{x^2}{r^2}\right] \qquad (32)$$

$$\mathbf{u}_{z,cl}(\mathbf{R}) = 0$$

where $r = \sqrt{x^2 + y^2}$.

Similarly, for a straight screw dislocation, we have

$$\mathbf{F}_x(\mathbf{k}_\parallel) = 0 = \mathbf{F}_y(\mathbf{k}_\parallel)$$

$$\mathbf{F}_z(\mathbf{k}_\parallel) = F_z(\mathbf{s}) = \frac{b}{k_\parallel^2}\frac{k_y}{k_x} \qquad (33)$$

then after the Fourier transform Eq. (16), we have

$$\mathbf{u}_{x,cl}(\mathbf{R}) = 0 = \mathbf{u}_{y,cl}(\mathbf{R})$$

$$\mathbf{u}_{z,cl}(\mathbf{R}) = \frac{b}{2\pi}\arctan\left(\frac{y}{x}\right) \qquad (34)$$

Both Eq. (32) for an edge dislocation and Eq. (34) for a screw dislocation are well-known classical results, validating the reducibility of mode expansion procedure.

For more complex systems, such as anisotropic materials, we could start from the dynamic atomic displacement field $\mathbf{u}_{cl}(\mathbf{R})$ and perform an inverse Fourier transform in Eq. (14) to find out the particular form of $\mathbf{F(k)}$.

### 3.7 The dislon formalism: *What is* $\mathbf{k}$?

Before the second quantization, we would like to clarify the choice of the wavevector $\mathbf{k}$. Given the broken translational symmetry caused by a dislocation, it seems unlikely for $\mathbf{k}$ to lie in the first Brillouin zone of the original periodic lattice at first glance. In this sense, any quasi-continuous $\mathbf{k}$ allowing for the substitution

$$\sum_k = \frac{L^3}{(2\pi)^3}\int d^3\mathbf{k}$$, with integral range $\mathbf{k}_i \in (-\infty, +\infty)$

will allow for the transformation of $\mathbf{F(k_\parallel)}$ to classical dislocation's displacement, and thereby become a valid choice of $\mathbf{k}$ values. For instance, the traveling-wave $\mathbf{k}$- values along the $i^{\text{th}}$ ($i$=1,2,3) direction of the box with length $L$, i.e.

$$\mathbf{k}_i = \frac{2\pi n_i}{L}, n = 0, \pm 1, \pm 2, \dots \quad (35)$$

are certainly quasi-continuous, hence can be considered as a valid $\mathbf{k}$-set choice if we only want to define a dislocation as in Eq. (14) with valid Fourier transform that is reducible to a classical dislocation's displacement.

However, the freedom of an arbitrary choice of $\mathbf{k}$ disappears when we study the electron-dislon interaction or phonon-dislon interaction. For electrons in a crystalline solid, assuming $N$ atoms in one direction with lattice constant $a$, with $L=Na$, the corresponding crystal momentum $\mathbf{k}$ along the $i^{\text{th}}$ ($i$=1,2,3) direction can be written as

$$\mathbf{k}_i = \frac{\pi n_i}{L}, n_i = -N, -N+1, \dots, N-2, N-1 \quad (36)$$

with the range $\mathbf{k}_i \in \left[ -\pi/a, +\pi/a \right)$. In other words, the crystalline solid with lattice constant $a$ will set a lower and upper limit of momentum and makes the displacement field expressions valid approximately.

To remedy this problem, we notice that the displacement field expressions, like Eqs. (32) and (34), are valid rigorously only in the continuous limit. If we plan to obtain a low-energy, long-wavelength effective field theory for a continuous elastic medium, we could set the lattice parameter $a \rightarrow 0$ in Eq. (36). In doing so, the resulting $\mathbf{k}$ for a dislocation would exactly match that of electrons and phonons, hence facilitates all calculation. Put differently, we do not need to construct a separately set of $\mathbf{k}$ for dislocation's mode expansion; the $\mathbf{k}$ values used for phonons and electrons, whether in the first Brillouin zone or in a continuous medium, are sufficient to define the mode expansion of dislocation.

Although the $\mathbf{k}$ magnitudes for electrons/phonons and dislocations can be chosen to be the same, they have a different interpretation. For electrons and phonons, with periodicity, $\mathbf{k}$ can be considered as both a generator of translational motion and good quantum number to label the quantum states of different translational states. On the other hand, for dislocations, without periodicity, $\mathbf{k}$ is no longer a good quantum number for translational motion, but can still be used to label quantum states by maintaining the mathematical rigor.

### 3.8 The dislon formalism: *From first to second quantization*

With the first-quantized dislocation field established, we are ready to construct the corresponding second-quantized dislocation field, aka the dislon. To understand

the procedure, we recall the second quantization of phonons. For a first-quantized phonon Hamiltonian

$$H_{ph} = \sum_{\mathbf{k}} \left( \frac{p_{\mathbf{k}}^{ph} p_{-\mathbf{k}}^{ph}}{2m} + \frac{1}{2} m \omega_{\mathbf{k}}^2 u_{\mathbf{k}}^{ph} u_{-\mathbf{k}}^{ph} \right) \quad (37)$$

with canonical commutation relation

$$[u_{\mathbf{k}}^{ph}, p_{\mathbf{q}}^{ph}] = i\hbar \delta_{\mathbf{kq}} \quad (38)$$

Then, defining the phonon annihilation and creation operator $b_{\mathbf{k}}$ and $b_{\mathbf{k}}^+$,

$$\begin{aligned} u_{\mathbf{k}}^{ph} &= \sqrt{\frac{\hbar}{2m\omega_{\mathbf{k}}}} \left( b_{\mathbf{k}} + b_{-\mathbf{k}}^+ \right) \\ p_{\mathbf{k}}^{ph} &= i\sqrt{\frac{\hbar m \omega_{\mathbf{k}}}{2}} \left( b_{-\mathbf{k}}^+ - b_{\mathbf{k}} \right) \end{aligned} \quad (39)$$

where $m$ is the atomic mass, and $\omega_{\mathbf{k}}$ is the phonon dispersion. Then, Eq. (37) is written as second-quantized form as

$$H_{ph} = \sum_{\mathbf{k}} \hbar \omega_{\mathbf{k}} \left( b_{\mathbf{k}}^+ b_{\mathbf{k}} + \frac{1}{2} \right) \quad (40)$$

with Bosonic commutation relation

$$[b_{\mathbf{k}}, b_{\mathbf{q}}^+] = \delta_{\mathbf{kq}} \quad (41)$$

Here, for the dislon case, we need to first find out the classical first-quantized Hamiltonian for a dislocation. This has been carried out in detail [198]. Briefly speaking, for a displacement field $\mathbf{u}(\mathbf{R})$, we will have the kinetic energy $T$:

$$T = \frac{\rho}{2} \int \sum_{i=1}^{3} \dot{\mathbf{u}}_i^2(\mathbf{R}) d^3\mathbf{R} \quad (42)$$

where $\rho$ is the density of the material. And the potential energy $U$:

$$U = \frac{1}{2} \int c_{ijkl} u_{ij} u_{kl} d^3\mathbf{R} \quad (43)$$

where $c_{ijkl}$ is the rank-4 stiffness tensor and $u_{ij}$ is the strain tensor. In an isotropic material, $c_{ijkl} \equiv \lambda \delta_{ij} \delta_{kl} + \mu(\delta_{ik}\delta_{jl} + \delta_{il}\delta_{jk})$. Using Eq. (30), the total dislocation Hamiltonian $H_D$ at the first-quantization level can be written as

$$H_D = T + U = \frac{1}{2L} \sum_{\mathbf{k}} T_{\mathbf{k}} \dot{u}_{\mathbf{k}} \dot{u}_{-\mathbf{k}} + \frac{1}{2L} \sum_{\mathbf{k}} W_{\mathbf{k}} u_{\mathbf{k}} u_{-\mathbf{k}} \quad (44)$$

where $T_{\mathbf{k}} \equiv \rho |F(\mathbf{k})|^2$ and $W_{\mathbf{k}} \equiv (\lambda + \mu)[\mathbf{k} \cdot \mathbf{F}(\mathbf{k})]^2 + \mu k^2 |F(\mathbf{k})|^2$. Then, defining

$$\begin{cases} u_{\mathbf{k}} = Z_{\mathbf{k}} \left[ a_{\mathbf{k}} + a_{-\mathbf{k}}^+ \right] \\ p_{\mathbf{k}} = \dfrac{i\hbar}{2Z_{\mathbf{k}}} \left[ a_{\mathbf{k}}^+ - a_{-\mathbf{k}} \right] \end{cases} \qquad (45)$$

with $Z_{\mathbf{k}} \equiv \sqrt{\hbar / 2 m_{\mathbf{k}} \Omega_{\mathbf{k}}}$, $m_{\mathbf{k}} \equiv T_{\mathbf{k}} / L$ and $\Omega_{\mathbf{k}} = \sqrt{W_{\mathbf{k}} / T_{\mathbf{k}}}$. It seems that the dislon Hamiltonian in the first-quantized form Eq. (44) can be written in second-quantized form, as analogously to the phonon case shown in Eq. (40).

### 3.9 Breakdown of canonical quantization and the quasi-Bosonic behavior of dislon

However, unlike the standard canonical quantization procedure for non-interacting particles, here, we need to check the consistency between the boundary condition and the canonical quantization. It is well known that when a constraint exists, the canonical quantization procedure is allowed to break down [202]. As an example, for a single free quantum particle, we have the canonical quantization condition:

$$[x, p_x] = i\hbar, \ [x, p_y] = i\hbar$$
$$[y, p_y] = i\hbar, \ [y, p_x] = i\hbar \qquad (46)$$

If we restrain the particle motion along the line $x + y = 0$, then we should have $[x + y, p_x] = [0, p_x] = 0$. However, from the canonical quantization condition, we should have

$$0 = [x + y, p_x] = [x, p_x] + [y, p_x] = i\hbar$$

resulting in apparent contradiction. For the current dislocation case, if we impose the canonical quantization condition

$$[u_{\mathbf{k}}, p_{-\mathbf{k}}] = i\hbar \qquad (47)x$$

then using the boundary condition Eq. (28), we have

$$i\hbar = \lim_{k_z \to 0} i\hbar = \lim_{k_z \to 0} \left( u_{\mathbf{k}} p_{-\mathbf{k}} - p_{-\mathbf{k}} u_{\mathbf{k}} \right)$$
$$= \lim_{k_z \to 0} \left( p_{-\mathbf{k}} - p_{-\mathbf{k}} \right) = 0 \qquad (48)$$

also leading to apparent contraction. This contradiction indicates a breakdown of canonical quantization condition Eqs. (20) and (47) (labeled with "x" as invalid) due to the constraint of the dislocation Eq. (28).

To remedy the inconsistency, one approach is to implement the Dirac's canonical quantization [203], which is cumbersome by introducing additional auxiliary variables. Fortunately, a much simpler solution has been found [198]. Instead of imposing Eq. (47), which is equivalent to $[a_{\mathbf{k}}, a_{\mathbf{q}}^+] = \delta_{\mathbf{kq}}$, a different commutation condition:

$$[a_{\mathbf{k}}, a_{\mathbf{q}}^+] = \delta_{\mathbf{kq}} \, \text{sgn}(\mathbf{k}) \qquad (49)$$

is adopted, where sgn is the vector sign function. Then, for any $\mathbf{k} \neq \mathbf{q}$, Eq. (49) satisfies conventional Bosonic statistics. For $\mathbf{k} = \mathbf{q}$ but $\text{sgn}(\mathbf{k}) > 0$, Eq. (49) is still valid for conventional Bosons. The only difference occurs when $\text{sgn}(\mathbf{k}) < 0$. To reduce the Eq. (49) to a more familiar form, the concept of supersymmetric Boson sea [204,205] was introduced [58,198], i.e.

$$a_{\mathbf{k}1} \equiv a_{\mathbf{k}}, \text{ when } \text{sgn}(\mathbf{k}) > 0$$
$$a_{\mathbf{k}2}^+ \equiv a_{-\mathbf{k}}, \text{ when } \text{sgn}(\mathbf{k}) < 0 \qquad (50)$$

Then both $a_{\mathbf{k}1}$ and $a_{\mathbf{k}2}$ satisfy Bosonic statistics, and will lead to consistency with the constraint Eq. (28). Substituting Eq. (50) into Eq. (45) and further using Eq. (30), the displacement field of dislocation at the second-quantization level can be written as

$$\mathbf{u}(\mathbf{R}) = \frac{1}{L^{3/2}} \sum_{\text{sgn}\,\mathbf{k} > 0} e^{i\mathbf{k} \cdot \mathbf{R}} \mathbf{F}(\mathbf{k}) \sqrt{\frac{\hbar}{2T_{\mathbf{k}} \Omega_{\mathbf{k}}}} \left[ a_{\mathbf{k}1} + a_{\mathbf{k}2} \right] \qquad (51)$$

with boundary condition

$$\lim_{k_z \to 0} \left( a_{\mathbf{k}1} + a_{\mathbf{k}2} \right) = \lim_{k_z \to 0} \sqrt{\frac{2m_{\mathbf{k}} \Omega_{\mathbf{k}}}{\hbar}} \qquad (52)$$

What is worth mentioning here is that Eq. (52) depends on the size $L$ of the system since $m_{\mathbf{k}} = T_{\mathbf{k}} / L$. For one single dislocation lying inside an area $L^2$, the larger $L$ indicates a smaller weight of the dislocation; when $L \to \infty$, the classical dislocation's weight becomes negligible compared to the role of quantum fluctuation, reducing to normal phonons.

### 3.10 The Dislon Hamiltonian

One major advantage of the second quantization formalism is to simplify the dislon Hamiltonian. Substituting Eq. (50) into Eq. (45) and then using the fact that $p_{\mathbf{k}} = m_{\mathbf{k}} \dot{u}_{\mathbf{k}}$, the dislon Hamiltonian Eq. (44) can finally be written as

$$H_D = \sum_{\text{sgn}\,\mathbf{k} > 0} \hbar \Omega(\mathbf{k}) \left[ \left( a_{\mathbf{k}1}^+ a_{\mathbf{k}1} + \frac{1}{2} \right) + \left( a_{\mathbf{k}2}^+ a_{\mathbf{k}2} + \frac{1}{2} \right) \right] \qquad (53)$$

with boundary condition Eq. (52) valid. In other words, to quantize a dislocation and satisfy the topological constraint, instead of one Bosonic field, at least two Bosonic fields have to be introduced. This might be intuitively comparable to a Dirac monopole, which is also a topological defect described by two fields. Unlike a dislon which is described by two quantum fields $a_{\mathbf{k}1}$ and $a_{\mathbf{k}2}$, a Dirac monopole is described by two classical

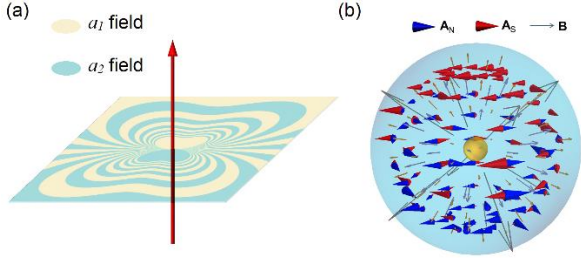

**Figure 17.** Comparison between a dislon (a) and a Dirac monopole (b). Both are topological defects described by two fields, but quantum and classical fields, respectively.

fields of magnetic vector potential $A_N$ and $A_S$ (Figure 17).

With dislon Hamiltonian Eq. (53) in hand, whenever we want to introduce dislocations into the existing electronic or phononic system, we can simply add the Hamiltonian Eq. (53), and then find the interaction mechanism which is related to the dislon displacement Eq. (51).

### 3.11 Straightforward Generalization: Anisotropy, Coulomb generalization, and vector quantization

Up to now, we have developed a comprehensive procedure to quantize a straight dislocation line with generic dislocation type, under the following premises: a) Isotropic continuous medium b) Dynamic vibrational effect taken into account c) Scalar quantization where the fluctuation along different directions are dependent, and d) Supersymmetric Boson sea to treat the topological invariant. As we will see shortly, all these premises can be relaxed to form a more general quantized dislocation theory.

For materials with anisotropy, the potential energy Eq. (43) is still valid, yet the rank-4 stiffness tensor $c_{ijkl}$ can be generalized to more complicated form for anisotropic solids [206]. However, a more complicated effective stiffness tensor can also be defined for surface, interface and nanostructured systems beyond bulk anisotropy, such as thin films, thin wires and small spherical particles [207]. For instance, for a spherical nanoparticle system with radius $R_0$, instead of using Eq. (43), the bulk potential energy density $U_{bulk}$ can be defined as

$$U_{bulk} = \frac{1}{2} C_{ijkl} u_{ij} u_{kl} + \frac{1}{6} C^{(3)}_{ijklmn} u_{ij} u_{kl} u_{mn} \quad (54)$$

where $C^{(3)}_{ijklmn}$ is the six-order elasticity tensor. Similarly, surface potential energy density $U_{surface}$ can be written as

$$U_{surface} = \frac{\left( \tau_{ij} u_{ij} + \frac{1}{2} Q_{ijkl} u_{ij} u_{kl} + \frac{1}{6} Q^{(3)}_{ijklmn} u_{ij} u_{kl} u_{mn} \right)}{R_0} \quad (55)$$

Then the effective stiffness $\bar{C}_{ijkl}$ under a self-equilibrium condition with bulk-dominant modulus can be written as

$$\bar{C}_{ijkl} = C_{ijkl} + \frac{1}{R_0} \left( Q_{ijkl} - Q^{(3)}_{ijklmn} C^{-1}_{mnpq} \tau_{pq} \right) \quad (56)$$

As a result, the effective stiffness tensor enables the study of dislocations on finite and confined systems. This generalization greatly expands the power of the dislon theory to study the effect of dislocations from bulk systems to nanomaterials.

A different scheme of generalization is to incorporate the Coulomb interaction, or any material-specific core effect. This can be done by generalizing the form of the dynamic expansion coefficient $\mathbf{F}(\mathbf{k})$, which appears in Eq. (14) and is defined in Eq. (29). Tracing back to the original form of $\mathbf{F}(\mathbf{k})$ [73], we see that it originates from the dynamic vibrational effect of a dislocation, which dominates the phonon-dislocation scattering process. To incorporate the Coulomb interaction of a dislocation core, we recall the third definition of a dislocation (Section 1.1), which is derived from the force equilibrium condition using the theory of elasticity [6]:

$$\mathbf{f}_i = \frac{\partial \sigma_{ij}}{\partial \mathbf{R}_j} \quad (57)$$

where $\mathbf{f}_i$ is the body force. Then, the Coulomb interaction for a charged dislocation line can be introduced by coupling the electrostatic potential with the elastic displacement $\mathbf{u}(\mathbf{R})$. Such a coupled system has been studied extensively for electrodynamics in continuous media [208], which facilitates the generalization of dislocations in dielectric materials. As an example, in a simple isotropic material subject to deformation under the presence of a uniform electric field $\mathbf{E}$, the change of electrostatic potential $\delta\phi$ can be written as

$$\delta\phi = \phi(\mathbf{R} - \mathbf{u}) - \phi(\mathbf{R}) = -\mathbf{u} \cdot \nabla\phi = \mathbf{u} \cdot \mathbf{E} \quad (58)$$

Then, the corresponding stress tensor can be written as

$$\sigma_{ij} = \sigma^0_{ij} - \frac{E^2}{8\pi} \left( \varepsilon - \rho \frac{\partial \varepsilon}{\partial \rho} \Big|_T \right) \delta_{ij} + \frac{\varepsilon E_i E_j}{4\pi} \quad (59)$$

in which $\sigma^0_{ij}$ is the portion without electric field, $\varepsilon$ and $\rho$ are the dielectric constant and density, respectively. The case of a dislocation generalizes Eq. (58) to a spatially-dependent $\mathbf{E}(\mathbf{R})$, but still satisfies a constitutive relation and thus forms a new equilibrium condition. By introducing this generalized force equilibrium condition with material-dependent electrostatic potential $\phi$, this approach enables the study of arbitrary dislocation core effects, such as different scattering rates due to core polymorphism [209].

Besides generalizing $c_{ijkl}$ and the form of $\mathbf{F(k)}$, another possibility is to relax Eq. (27), making the quantum fluctuation along each direction independent, or satisfying different conditions. This generalization procedure can be done by solving the corresponding dynamical matrix, just like the case of phonons. However, as we shall see shortly, if it comes to this complexity, alternative quantization approaches do exist.

### 3.12 Further Generalization: Lattice quantization and gauge quantization

The current quantization approach (Sections 3.2-3.10) is rigorous in continuous elastic media, where a dislocation arises as a singularity. If we are interested in a lattice system, the displacement field on each lattice site can be used, with the singularity eliminated naturally. This can be done using the Frank-Kontorova model [210-212], the Peierls model [213], or more realistic models [1]. For instance, for one edge dislocation in a simple cubic system (Figure 18), the corresponding lattice displacement field $\mathbf{u}(x, y)$ with misfit $\gamma$ can be written as [214]

$$
\begin{aligned}
\mathbf{u}_x(x, y) &= \frac{1}{2\mu}\left[\gamma Y - 2(1-\nu)\right]e^{-\gamma Y}\sin\gamma x \\
\mathbf{u}_y(x, y) &= \frac{1}{2\mu}\left\{\begin{array}{l}\left[\gamma Y - 1 + 2(1-\nu)\right]e^{-\gamma Y}\cos\gamma x \\ +1 - 2(1-\nu)\end{array}\right\}
\end{aligned}
\tag{60}
$$

where $Y = y - \dfrac{a}{2}$, with $a$ the lattice constant.

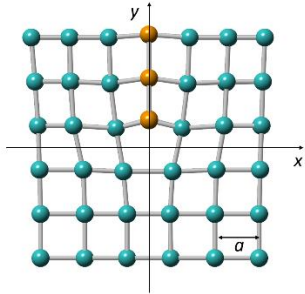

**Figure 18.** Coordinate system used for dislocation's lattice displacement.

As for screw dislocation, Maradudin derived the vertical lattice displacement in the row index $(m, n)$ [215]

$$
\mathbf{u}_z(m, n) = -\frac{b}{\pi^2}\iint\int_0^{\pi/2}dxdy\frac{\sin y}{\sin x}\frac{\sin 2mx\sin 2ny}{\sin^2 x + \sin^2 y}
\tag{61}
$$

In another example, using the force equilibrium condition and the lattice Green's function method, the displacement field of a dislocation in a 2D triangular crystal has also been derived [216,217]. In either case, a lattice displacement field may lead to a lattice quantization, which can be applied to examine any novel electronic, phononic and photonic modes near the dislocation core.

Beyond all previous examples, one additional possibility for the dislocation's quantization is to generalize Eqs. (12) - (14). The static solution of Eq. (14), such as Eq. (16), certainly satisfies the dislocation's definition Eq. (1), which is seen in Eqs. (31)-(34). However, it is worth mentioning that the same procedure also breaks a type of gauge symmetry, which leaves the dislocation Lagrangian unchanged under a set of local gauge transformations. Classical dislocations can be described by the classical affine gauge theory [218-221]. In this situation, the interaction naturally emerges from the minimal derivative coupling procedure with the matter Lagrangian. If properly quantized, a dislocation may behave as a gauge Boson beyond the harmonic-oscillator-like operator as in Eq. (49), which resembles the relation between quantum electrodynamics (QED) with U(1) gauge symmetry and the simplified scalar QED [222].

If, on the other hand, we restrict the quantization in 2D, a very recently developed approach based on fracton [223,224], which arises from the tensor-gauge theory – 2D quantum crystal may be adopted [225,226].

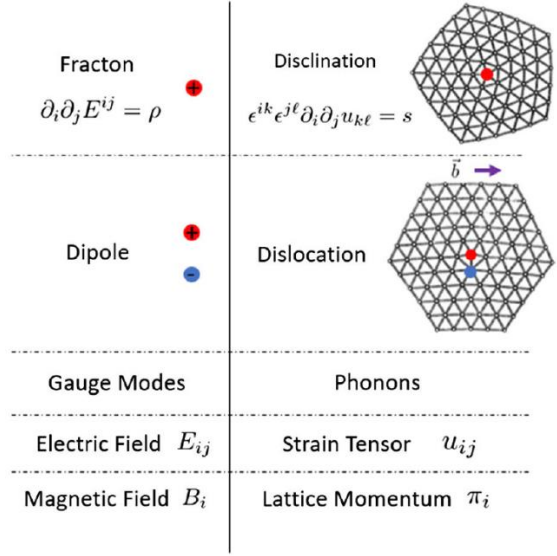

**Figure 19.** The correspondence between the fracton excitation and the lattice defects in 2D crystals. Figure adapted from [226].

### 3.13 Workflow of a general quantized dislocation problem

We recap the quantization process by providing a step-by-step flowchart to illustrate the general procedure to apply the dislon theory for a given dislocation related functionality problem (Figure 20).

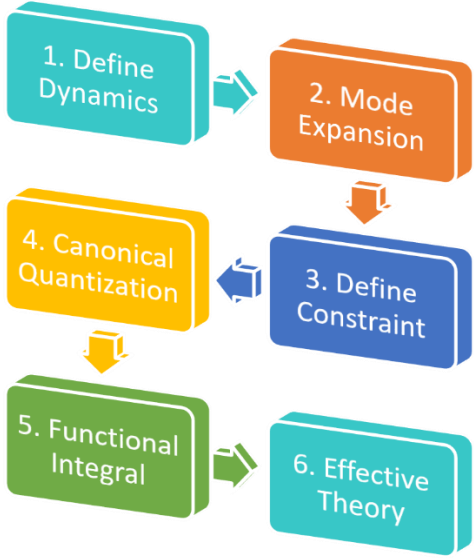

**Figure 20.** General workflow to apply the dislon theory.

*Step 1. Define dynamics.* We first need to identify the relevant dynamical interaction processes that come into play. Any physical process beyond classical strain field may fall this category. For instance, if we want to study phonon-dislocation scattering, the dynamic fluttering dislocation shall be considered; if we study possible states near a ferromagnetic dislocation as discussed in Section 2.4, then a spin index shall be introduced.

*Step 2. Mode expansion.* After identifying the relevant dynamic process, we may turn to a local force equilibrium condition by writing down the corresponding constitutive relation. For instance, when Coulomb, dynamic, and strain field co-exist, a process outlined in Section 3.11 can be adopted, resulting in a general mode expansion of a dynamic dislocation. A simple example is demonstrated in Section 3.2.

*Step 3. Define constraint.* We will also need to identify the corresponding constraint that enables a full reduction to a classical, static dislocation.

*Step 4. Canonical quantization.* By identifying the canonical coordinate and conjugate momentum, it is always possible to promote the classical dynamical variable to quantized operators, with the constraint serves as a boundary condition. The example procedure is shown in Sections 3.3 – 3.10.

*Step 5. Functional integral.* Due to the constrained nature of dynamics (Step 3), it is much more convenient to apply a functional integral approach instead of a Hamiltonian approach. Briefly speaking, if we have a constraint functional $C$ for dislocation's $a_1$ and $a_2$ fields:

$$C[a_1, a_2] = 0 \qquad (62)$$

Then, in the functional integral formalism, we can readily handle the constraint by adding a Dirac $\delta$-function $\delta(C[a_1, a_2])$ in the functional integral of dislon fields $\int Da_1 Da_2$.

*Step 6. Effective theory.* In functional integral form, the dislocation Hamiltonian $H_D$, interaction Hamiltonian $H_I$ and Hamiltonian of interest $H_0$ (whether electron, phonon or photon) is rewritten in terms of actions forms, $S_D$, $S_I$, and $S_0$, respectively. Then one way to write down the effective action $S_{\text{eff}}$ to see how dislocations effectively change the original system $S_0$ is given by

$$e^{-S_{\text{eff}}} = e^{-S_0} \int Da_1 Da_2 \delta(C[a_1, a_2]) e^{-S_D - S_I} \qquad (63)$$

With an effective action $S_{\text{eff}}$ in hand, we will then be able to compute almost all physical properties, at a microscopic level.

### 3.14 Advantages of dislon over empirical models

Before introducing the interaction problems, we summarize a few potential advantages of adopting the dislon theory to study materials functionalities influenced by dislocations.

*Simplicity*: Although sounds astonishing, the dislon theory has great structural simplicity. First, after going through the quantization procedure, the dislon Hamiltonian Eq. (53) closely resembles a simple harmonic oscillator with an extra constraint. Even so, it is sufficient to describe a generic type of line dislocation. Second, the quadratic form Eq. (53) greatly facilitates the derivation of an effective action Eq. (63) by a Gaussian integral. Third, for any problems associated with dislocations, we can simply add the dislocation Hamiltonian into the existing system, without the need to perform a case-by-case manipulation of the existing degrees of freedom.

*Capability*: As a quantum field theory, the dislon theory is capable of studying complex interacting systems, such as electron-electron correlation, electron-phonon interaction, electron-impurity interaction, etc., to arbitrarily high order. Such capability is essential to capture any qualitative change that is missed by first-order scattering theory. For instance, Anderson localization can be obtained when considering the maximally crossed Feynman diagrams self-consistently [227,228]. Therefore, the proper treatment of interactions may lead to full predictive power, beyond empirical models which often have a pre-defined physical scenario that is destined to happen.

*Generality*: On the one hand, the generality lies in the standardized procedure to incorporate dislocation, by simply adding the dislocation Hamiltonian $H_D$ and the interaction Hamiltonian $H_I$. As a result, it provides a powerful approach to compare with experiments. For any physical quantity $P$ changed by dislocations, almost all experiments perform control experiments without dislocations. Using dislon theory, both the main experiment with dislocation $P[H_{tot}] \equiv P[H_0 + H_I + H_D]$ and the latter control experiment $P[H_0]$ can be calculated; the ratio $P[H_{tot}]/P[H_0]$ can always be used to compare with experiments, even if we do not have a clear understanding of $H_0$ itself to allow for a quantitative comparison of $P$, directly. On the other hand, the generality lies in the procedure to compute physical properties. This includes at least thermodynamic properties such as specific heat, transport properties such as electronic and lattice thermal conductivity, magnetic properties such as permeability, optical properties such as dielectric function and optical absorption, and phase properties such as magnetic and superconducting transitions, where standard procedures to compute these properties from a Hamiltonian are all well established.

We take heat capacity as an example. Atomistic simulation shows that the temperature dependence of lattice heat capacity is complicated even in simple metals, with peaks, valleys and dislocation-type dependence. (Figure 21) [229]. The dislon theory offers a viable approach to directly compute the heat capacity with all factors considered; thus, it may provide insight to understand the origin of these profiles without going through large-scale computation.

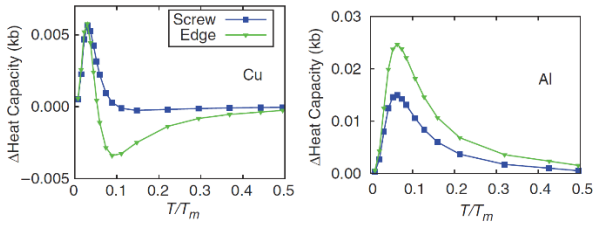

**Figure 21.** Relative heat capacity change in simple metals as a function of temperature. Adapted from [229].

## 4 Electron-Dislon Interaction

With the dislon Hamiltonian in hand, we are in good shape to study interaction problems. Since the formalism has been provided in detail in recent dislon studies [57,58,198,230,231], starting from this section, we keep the formalism semi-quantitative to emphasize the new phenomena.

### 4.1 Dislon excitation

After quantization, the dislon becomes a quantized particle with its own excitation spectra. Given the lattice displacement nature of dislocation, such an excitation shares the same energy scale of a phonon (Figure 22).

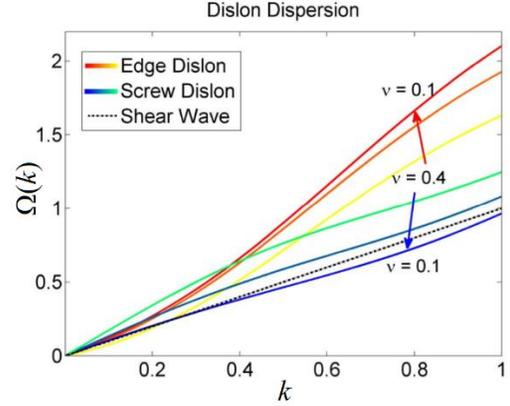

**Figure 22.** The dislon excitation spectra for edge and screw dislocations in an isotropic medium. Such excitation can be considered as local vibrational modes along the dislocation line. Figure adapted from [57].

In fact, with shear velocity $v_s$, the dislon dispersion along the dislocation line $\Omega(k)$ can be written as [57]

$$\Omega(k) = v_s k \times \eta(k) \tag{64}$$

where $\eta(k)$ is a $k$-dependent factor that accounts for the quantum correction; $\eta(k) = 1$ is the conventional linear transverse phonon dispersion. Such an excitation decays away from the dislocation core and has closed analytical form for both edge and screw dislocations in isotropic materials. Moreover, the comparable energy scale between a dislon and a phonon, together with the ubiquitous nature of electron-phonon coupling, strongly suggests the role that a dynamical dislocation may play in electronic properties, which has long been overlooked. Such dynamical processes may also be helpful to explain the resonance electron – line-defect scattering in metals which was "*still very much an open question*" since the 1970s [44].

### 4.2 Single-electron energy oscillation: a new type of quantum oscillation

One unexpected outcome that arises from the dynamic dislocation is the oscillatory electron energy near a dislocation core. The electron-ion interaction energy density $\varepsilon_{e-ion}$ near an atom displaced from $\mathbf{R_0}$ to $\mathbf{R}$ can be written as

$$\varepsilon_{e-ion} = \rho_e(\mathbf{R})\nabla V(\mathbf{R} - \mathbf{R_0}) \cdot \mathbf{u}(\mathbf{R_0}) \tag{65}$$

which describes the interaction between the electron density $\rho_e(\mathbf{R})$ and a nucleus through the Coulomb deformation potential $\nabla V$ caused by the dislocation displacement $\mathbf{u}$. From this interaction, the electron self-energy can be computed.

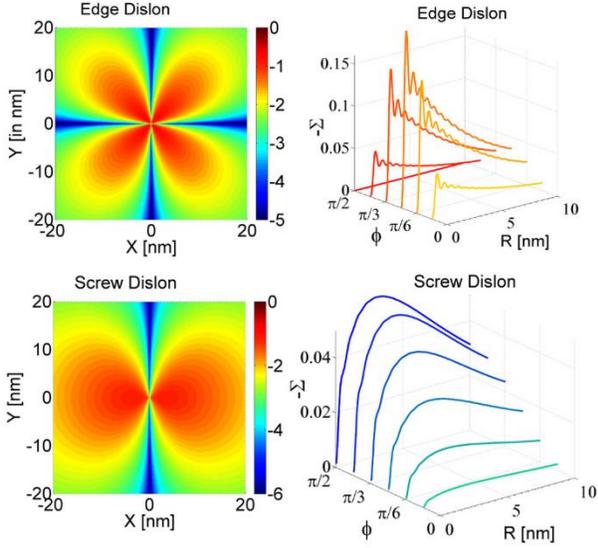

**Figure 23.** The electron energy shift as a function of spatial location, for edge (upper) and screw (lower) dislon. The left figures are plotted in logarithmic scale, while the right figures are the line cuts of the left figures in polar coordinate. Figures are adapted from [57].

As a result, the electron energy develops an anisotropic oscillatory pattern when the electron is moving near the dislocation core (Figure 23). Such an oscillation is stronger in edge dislocations than screw dislocations, due to a longitudinal dilation/compression of the unit cell in edge dislocations; screw dislocation, on the other hand, contains pure shear strain in isotropic material and thus does not contribute to this oscillation [56]. Compared to Friedel oscillation near a charged impurity, which is the electron density oscillation in Fermi liquids, this type of electron energy oscillation can be sustained even with a single electron present.

### 4.3 Predicting critical temperature in dislocated superconductors

Another dislon application involves the computation of the superconducting transition temperature $T_c$ with dislocations. Following the workflow described in Section 3.13 and using the electron-ion deformation potential Eq. (65), the final electron effective Hamiltonian $H_{\text{eff}}$ can be written as [58]

$$H_{\text{eff}} = H_0 + H_{cl} + H_{qu} \qquad (66)$$

where $H_0$ is the free electron Hamiltonian. In the second-quantized form, defining electron creation and annihilation $c_{\mathbf{k}\sigma}^+$ and $c_{\mathbf{k}\sigma}$, we have

$$H_0 = \sum_{\mathbf{k}\sigma} (\varepsilon_{\mathbf{k}} - \mu) c_{\mathbf{k}\sigma}^+ c_{\mathbf{k}\sigma} \qquad (67)$$

The classical electron-dislon scattering Hamiltonian $H_{cl}$ is written as:

$$H_{cl} = \sum_{\mathbf{k}\sigma} \sum_{\mathbf{s}} A_{\mathbf{s}} c_{\mathbf{k}+\mathbf{s}\sigma}^+ c_{\mathbf{k}\sigma} \qquad (68)$$

This equation describes the electron scattering processes $\mathbf{k} \rightarrow \mathbf{k} + \mathbf{s}$ by a dislocation ( Figure 24a, straight lines) in momentum space, with scattering amplitude $A_{\mathbf{s}}$. Here, $\mathbf{s}$ is the momentum change only perpendicular to the dislocation line direction.

The quantum Hamiltonian $H_{qu}$ is written as

$$H_{qu} = -g_D \sum_{\mathbf{q}\mathbf{k}\mathbf{k}'} c_{\mathbf{k}+\mathbf{q}\uparrow}^+ c_{-\mathbf{k}\downarrow}^+ c_{-\mathbf{k}'+\mathbf{q}\downarrow} c_{\mathbf{k}'\uparrow} \qquad (69)$$

which describes the attraction between electrons, which is exactly the BCS mechanism, but with a different coupling constant $g_D$ ( Figure 24a, wavy lines).

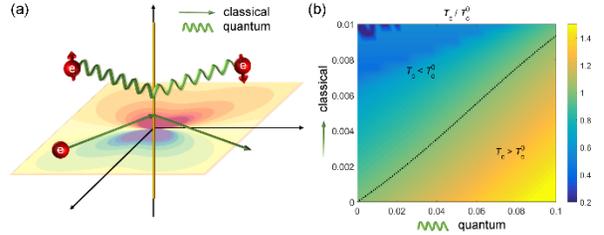

**Figure 24.** (a) Two types of competing electron-dislocation interactions derived from the dislon theory.

Here we want to emphasize that Eqs. (66)-(69) are not artificially added or intuitively guessed; they are rigorously derived by solely using the dislon Hamiltonian Eq. (53) and the boundary condition Eq. (52), followed by the functional integral procedure (Section 3.13 and Figure 20), without any other assumption. Even so, each term has clear physical interpretation. Most importantly, when derived in this way, both the classical scattering amplitude $A_{\mathbf{s}}$ and the quantum electron-dislon coupling strength $g_D$ are functions of materials electronic and mechanical parameters, enabling a "first-principle" determination of $T_c$. In particular, the Fourier transform of $A_{\mathbf{s}}$ leads to the dislocation scattering potential Eq. (6), which is indeed the approach to determine the constant $C$ in Eq. (9).

To proceed, we notice that the electron-dislocation relaxation rate $\Gamma_D$ from classical scattering Eq. (68) can be computed using Fermi's Golden rule:

$$\Gamma_D = \sum_{\mathbf{s}} \left| \langle \mathbf{k} + \mathbf{s} | A_{\mathbf{s}} | \mathbf{k} \rangle \right|^2 \delta(\varepsilon_{\mathbf{k}+\mathbf{s}} - \varepsilon_{\mathbf{k}}) \Big|_{\varepsilon_{\mathbf{k}} = \varepsilon_F} \qquad (70)$$

Then, in the end, a generalized BCS $T_c$ equation incorporating dislocations can be obtained [58]

$$\frac{1}{g_{ph} + g_D} = N(\mu) \int_0^{\theta_D'} d\xi \sum_{s=\pm 1} \tanh\left(\frac{\xi + is\Gamma_D}{2T_c}\right) \Big/ 2\xi \quad (71)$$

where $g_{ph}$ is the electron-phonon coupling constant, and $\theta_D'$ is the renormalized Debye frequency. With Eq. (71) at hand, it becomes possible to compute the transition temperature $T_c$. A simplified expression of $g_D$ and $\Gamma_D$ can be written as:

$$\Gamma_D = \frac{\pi m^*}{4\hbar^2 k_F^2 k_{TF}^4} \left(Ze^2 n\right) n_{dis} b^2 \left(\frac{1-2\nu}{1-\nu}\right)^2$$

$$g_D = \frac{n_{dis}}{L} \frac{\left(4\pi Ze^2 n\right)^2}{2k_{TF}^4 (\lambda + 2\mu)} \quad (72)$$

where $m^*$ is the effective mass, $k_F$ is Fermi wavevector, $k_{TF}$ is Thomas-Fermi screening wavevector, $n$ is atomic number density, $n_{dis}$ is dislocation density, $b$ is Burgers vector, $\nu$ is Poisson ratio, $\lambda$ and $\mu$ are Lame parameters, and $L$ is the box size whose explicit appearances is caused by different scaling between the classical scattering and quantum fluctuation. For Poisson ratio $\nu = 1/2$ (purely elastic, rubber-like medium), there is no classical scattering but quantum fluctuation is still present, which is reasonable.

Without dislocations, $g_D = 0 = \Gamma_D$, and Eq. (71) immediately reduces to the conventional BCS theory for computing the transition temperature $T_c^0$:

$$\frac{1}{g_{ph}} = N(\mu) \int_0^{\theta_D} d\xi \tanh\left(\frac{\xi}{2T_c^0}\right) \Big/ \xi \quad (73)$$

From Eq. (73), with an experimental $T_c^0$, we could obtain the $N(\mu)g_{ph}$ which can subsequently be substituted into Eq. (71).

Eq. (71) implies a competing mechanism between classical scattering $\Gamma_D$ and quantum coupling $g_D$, as shown in Figure 24b: When the classical term $\Gamma_D$ dominates, $T_c$ is reduced compared to the pristine material with $T_c^0$, and vice versa. A separation line with $T_c = T_c^0$ is seen clearly (black dashed line in Figure 24b), indicating an unchanged $T_c$ even with high dislocation density since both $g_D = 0 = \Gamma_D$. Since not all parameters are available for a given material, and the $T_c = T_c^0$ line is

a straight line, the ratio $g_D / \Gamma_D$ becomes a more useful indicator of the trend of $T_c$ change:

$$\frac{g_D / g_{ph}}{\Gamma_D / \theta_D} \propto \left(\frac{1-\nu}{1-2\nu}\right)^2 \frac{k_F^2}{m^* L(\lambda + 2\mu) b^2} \frac{\theta_D N(\mu)}{N(\mu)g_{ph}} \quad (74)$$

Then, with Eqs. (71) - (74) at hand, we are in good shape to compare with experiments, as shown in Table 2. Since $T_c^0$ from experiment is needed to acquire $N(\mu)g_{ph}$ as input for Eq. (71), it is only necessary to compute the ratio $T_c / T_c^0$:

**Table 2.** Experimentally measured and theoretically computed transition temperature ratio. Adapted from [58].

| $T_c/T_c^0$ | Al | In | Nb | Pb | Sn |
|---|---|---|---|---|---|
| Experiment | 1.00 | 1.04 | 1.06 | 1.00 | 1.06 |
| Theory | 0.94 | 1.07 | 1.06 | 1.03 | 1.06 |
| | Ta | Ti | Tl | V | Zn |
| Experiment | 1.00 | 0.75 | 1.13 | 1.09 | 1.54 |
| Theory | 1.01 | 0.87 | 1.07 | 1.05 | 1.28 |

Eq. (74) also provides a guideline engineering dislocations to enhance the $T_c$ in superconductors, through small effective mass, low elastic moduli and in a confined environment. This guideline may further be used to explain the potential dislocation-induced superconductivity in semiconductor superlattices [232-234], which fits all these requirements (Figure 25).

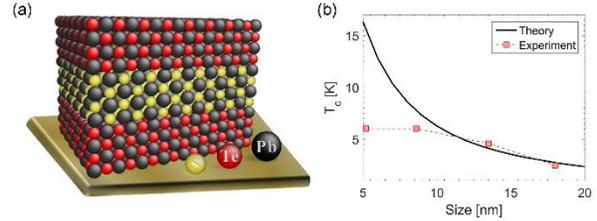

**Figure 25.** Emerging superconductivity in a semiconductor superlattice. Adapted from [58].

## 5 Phonon-Dislon interaction

Another primary application of the dislon theory lies in the phonon-dislocation interaction, dominated by the fluttering interaction, which is the drag-like coupling $\dot{\mathbf{u}}_{ph} \cdot \dot{\mathbf{u}}_D$ originated from the cross term in kinetic energy $\frac{1}{2}\rho\left(\dot{\mathbf{u}}_{ph} + \dot{\mathbf{u}}_D\right)^2$; the strong fluttering interaction leads to very different phenomenology compared to the electron-phonon interaction and can be wildly different from classical theories after quantization.

### 5.1 Non-perturbative phonon transport beyond the relaxation time approximation

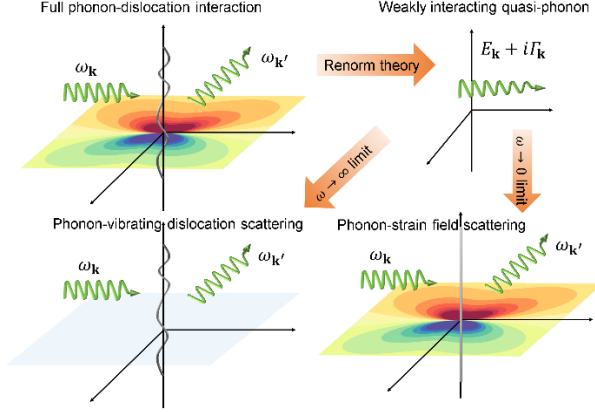

**Figure 26.** Quasi-phonon as a result of the phonon-dislocation interaction. The low-frequency and high-frequency limits of the quasi-phonon are the strain scattering and the dynamic fluttering scattering, respectively. Figure adapted from [231].

The primary result from dislon theory is that the $\dot{\mathbf{u}}_{ph} \cdot \dot{\mathbf{u}}_D$ fluttering interaction leads to an effective phonon theory beyond any perturbation theory. To see this, by following the workflow described in Section 3.13, the effective phonon action can be written as [231]

$$S_{\text{eff},ph} = \sum_k \frac{1}{2} \bar{\phi}_k \left( D_{0k} - J_k + \sqrt{D_{0k}^2 + J_k^2} \right) \phi_k \quad (75)$$

where $D_{0k}$ denotes a non-interacting phonon part, $\bar{\phi}$ and $\phi$ are the phonon fields, $J_k$ is the phonon-dislocation coupling strength, and $k$ is a summation of all modes (including all momenta and all Matsubara frequencies). When $J_k = 0$, Eq. (75) is reduced to a conventional free phonon theory,

$$S_{0,ph} = \sum_k \bar{\phi}_k D_{0k} \phi_k = \sum_k \bar{\phi}_k \left( -i\omega_n + \omega_{\mathbf{k}} \right) \phi_k \quad (76)$$

In the Matsubara frequency domain, $D_{0k} = -i\omega_n + \omega_{\mathbf{k}}$ is the inverse propagator, where $\omega_n$ is the Matsubara frequency and $\omega_{\mathbf{k}}$ is the phonon dispersion.

The situation beyond perturbation can be seen directly from Eq. (75). Since the coupling constant $J_k$ lies inside the "$\sqrt{\phantom{x}}$" operation, a Taylor expansion will naturally lead to an infinite order of $J_k$. This is in sharp contrast to perturbation theory, such as electron-dislocation effective theory Eq. (68) and the first-order perturbation Eq. (70). Through numerical procedure, we could approximate Eq. (75) in terms of a simpler theory that resembles Eq. (76),

$$S_{\text{eff},ph} = \sum_n \bar{\phi}_n \left( -i\omega_n + E_{\mathbf{k}} + i\Gamma_{\mathbf{k}} \right) \phi_n \quad (77)$$

i.e., a phonon interacting with a dislon (Eq. (75)) now behaves as a quasi-phonon, $\omega_{\mathbf{k}} \rightarrow E_{\mathbf{k}} + i\Gamma_{\mathbf{k}}$, where phonon energy $\omega_{\mathbf{k}}$ is changed to $E_{\mathbf{k}}$, with a finite lifetime as the imaginary part ( Figure 26).

Additionally, the debate discussed in Section 2.2, whether the phonon-dislocation interaction mechanism is static strain scattering or dynamic scattering, seems to be readily solved. By treating a dislocation as a quantum field, the long-range static part and the local dynamic part are simultaneously captured as the "field" and "quantum" aspect of the dislon, respectively. In particular, the fluttering and the static strain scattering can be considered as the high- and low- frequency limit, respectively ( Figure *26*), unifying the debate.

## 5.2 Energy shift of the quasi-phonon

After the phonon-dislocation interaction, a phonon $\omega_{\mathbf{k}}$ is renormalized as a quasi-phonon, with shifted phonon energy $E_{\mathbf{k}}$ and a finite relaxation rate $\Gamma_{\mathbf{k}}$. For $E_{\mathbf{k}}$, the real part of the quasi-phonon, the main predictions from the dislon theory is anisotropic phonon softening (yellow-dashed lines in Figure 27b). This can be verified through lattice dynamics simulations for a single dislocation in a hypothetical simple cubic crystal (Figure 27b), where the softening is seen clearly when compared with the pristine crystal (Figure 27a). An independent *ab initio* calculation of phonon-dislocation interaction in silicon with dislocation pairs (Figure 27c) shows a qualitatively identical feature of phonon softening (Figure 27d).

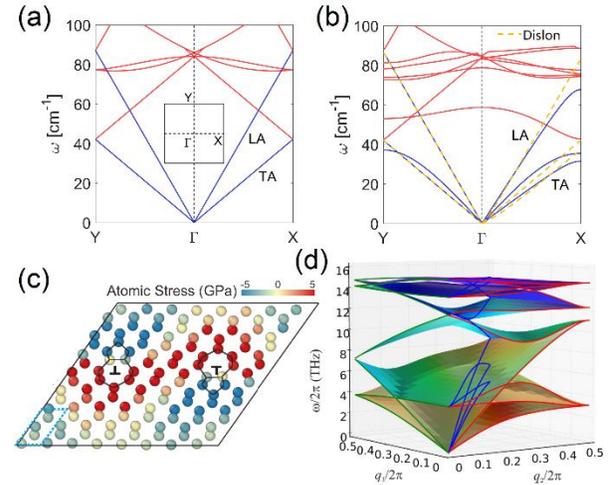

**Figure 27.** Phonon dispersion in a simple cubic crystal without (a) and with a dislocation (b), using lattice dynamics simulations and compared with dislon theory (yellow lines). Phonon dispersion relation (d) of dislocated Si (c) from *ab initio* calculations, showing a similar anisotropic softening feature. Figures are adapted from [231,235].

### 5.3 The relaxation time resonance

The effect on relaxation time is more striking. From classical theory, the dislocation scattering rates are monotonically dependent on frequency [236]:

$$\tau_{dyn}^{-1} \propto \omega^{-1}$$
$$\tau_{stat}^{-1} \propto \omega \qquad (78)$$
$$\tau_{core}^{-1} \propto \omega^3$$

where *dyn, stat, core* denote dynamic, static strain and dislocation core scattering, respectively. However, the dislon theory shows a clear resonance peak ( Figure *28*a) [231]. The resonance feature indicates that the phonon-dislocation scattering behaves as static at low frequency, but behaves as dynamic at high frequency, which is highly reasonable. Although the non-perturbative feature prevents an analytical expression of relaxation time, it can generally be inferred that

$$\tau_{ph-dis}^{-1} \propto \frac{\omega}{\Omega^2 + \omega^2} \qquad (79)$$

where $\Omega$ is the dislon dispersion.

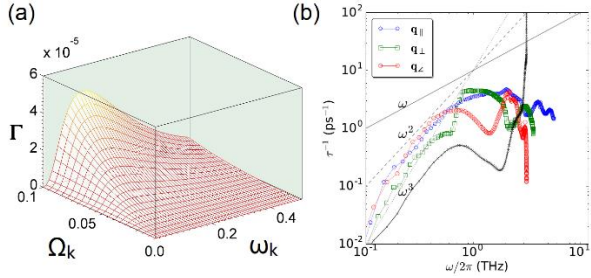

**Figure 28.** Phonon-dislocation relaxation rate from the dislon theory (a), and from *ab initio* calculations. In both cases, a non-monotonic, resonance-like behavior is shown, indicating a breakdown of all classical models. Figures are adapted from [231,235].

To verify this assertion, since the lattice dynamics simulation was performed in the harmonic approximation, it cannot be used to test the dislon theory. On the other hand, the *ab initio* phonon-dislocation calculation does shown a series of unusual phenomena, including the resonant-like feature in the relaxation rate, which is wildly different from Eq. (78). This concludes that "*Because of the breakdown of the Born approximation, earlier literature models fail, even qualitatively*", which is mentioned in [235].

### 6 Grand Journey with Dislon

Imagine the new opportunities brought by the quantization of phonons compared to classical lattice waves; the dislon may also bring other opportunities beyond static dislocation. Despite some initial success in electron- and phonon- dislocation scattering, we believe the dislon is still at its infant stage. In this Section, we introduce a few opportunities that may immediately be perceived using the current dislon framework.

#### 6.1 Dislon-induced quantum phase transition

A quantum phase transition driven by quantum fluctuation can happen when there are competing interactions, contrary to classical phase transition [237]. In light of this, the phonon-dislocation interaction composed of both the competing anharmonic interaction and the drag-like dynamic fluttering interaction [73,201] provides an ideal platform to achieve phonon quantum phase transition: the dynamic interaction action is $S_{dyn} \sim \int \dot{\mathbf{u}} \cdot \dot{\mathbf{u}}_{ph}$, while the anharmonic interaction action is $S_{anh} \sim \int \mathbf{u}\mathbf{u}_{ph}^2$. Therefore, an effective phonon action can be obtained following the workflow in Section 3.13, which shall contain both a quartic term $u_{ph}^4$ and a quadratic term $u_{ph}^2$, resembling Landau's theory for continuous phase transition, but for the phonon displacement field.

#### 6.2 Dislon-induced topological phase transition

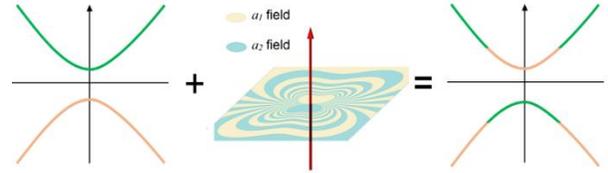

**Figure 29.** A band insulator without band inversion (left), when interacting with dislon fields (middle), may lead to a topological insulator (right) with band inversion due to the bandgap tuning effect of dislocations.

Many research efforts in topological materials have been directed toward new materials search. If there is one approach that directly turns a topologically trivial material into a topological insulator, it opens up vast opportunities to expand the family of topological insulators based on introducing dislocations to existing materials without the need to search for new materials. Since dislocations can induce large strain field [2] which further induce a substantial bandgap change [238-240], it is highly desirable to explore the dislocation-tuned bandgap and possible dislocation-induced topological phase transition in realistic materials, using the dislon theory (Figure 29).

Besides the strain field, the dynamic part may also tune the bandgap, just like a phonon does [241]. To achieve this, we start from a material described by a model Hamiltonian $H_0$, say a massless Dirac fermion with time-reversal-symmetry:

$$H_{2\times2}(\mathbf{k}) = -k_x\sigma_x - k_y\sigma_y \qquad (80)$$

with trivial topology $C = 0$, where $C$ is the Chern number. Given the interaction between electrons and dislocations $H_{e-dis}$, it can generate a self-energy matrix $\Sigma(\mathbf{k})$ and thereby effective mass whose sign change can drive the system to topologically nontrivial $C \neq 0$. Compared to the phonon-induced topological phase transition which needs sufficiently high temperature for sufficient phonon population [241], dislocations are more controllable through dislocation density.

### 6.3 Dislon-enhanced Anderson localization

Sufficient randomness can cause the disappearance of electron diffusion called Anderson localization [242]. In the past, Anderson localization has been studied in a wide variety of systems, such as disordered electronic systems [243,244], classical waves [245] and phonons [246], photons [247,248] and even Bose-Einstein condensates [249,250], in 3D, and lower dimensional 2D [251,252] and 1D systems [253,254]. In all these situations, Anderson localization can be considered as a strong coherent interference effect between multiple-scattering paths of waves in the presence of randomness. However, the direct observation of Anderson localization in bulk atomic crystals has long been challenging [255]; we need either stronger randomness or weaker electron scattering for realizing localization in realistic crystalline solids. In light of this, dislocations offer some great advantages to achieve Anderson localization compared with point impurities. On the one hand, dislocations can induce stronger randomness. Put another way, dislocations are 1D extended defects, hence the effective system dimension is reduced, and lower-dimension favors stronger Anderson localization. On the other hand, due to the long-range strain field, dislocations are further correlated in a highly dislocated material, in contrast to independent impurities. Since correlated disorders can enhance the effect of localization for both electrons and elastic waves [256,257], the correlation between dislocations offers an extra incentive to enhance localization. As a result, the possible Anderson localization driven by random dislocations in crystalline materials is worthwhile to explore, for both electrons and phonons.

### 7 Conclusions and Perspective

In this article, we have reviewed the recent progress on the theoretical effort to quantize a dislocation, aka the dislon theory. The unsatisfactory classical dislocation functionality theories and phenomenology leave plenty of hints to call for a quantization approach (Section 2). After quantization (Section 3), the interaction problems can be studied systematically by following the conventional quantum many-body theory (Sections 4-6). In retrospect, we may be wondering why the quantization procedure works in the first place. As an extended defect, a dislocation has finite spatial extent beyond point defect. The spatial extent endows a dislocation with internal dynamical structure, which is then captured by quantum fields. The co-existence of spatial extent ("field" part) and the resulting internal dynamics ("quantum" part) brings up the question, of whether any extended defect can (or shall) be described by quantum fields:

$$\text{Extended Defect} \overset{?}{=} \text{Quantum Field} \qquad (81)$$

The answer to the question Eq. (81) may stimulate further studies to understand the role that extended defects may play in materials quantum and functional properties, at a quantitative many-body level, far beyond empirical models.

### Acknowledgements

M.L. would thank the support from MIT faculty startup funding, and also thank G. Chen, M. S. Dresselhaus, and G. D. Mahan for their long-term support on this topic.